\documentclass[english,aps,amsmath,amssymb,reprint,twocolumn,pra,superscriptaddress,notitlepage,showpacs,tightenlines]{revtex4-1}
\usepackage[T1]{fontenc}
\usepackage[latin9]{inputenc}
\usepackage{babel}
\usepackage{bm}
\usepackage{amsmath}
\usepackage{amssymb}
\usepackage{graphicx}
\usepackage[unicode=true,pdfusetitle,
 bookmarks=true,bookmarksnumbered=false,bookmarksopen=false,
 breaklinks=false,pdfborder={0 0 1},backref=false,colorlinks=false]
 {hyperref}
\hypersetup{
 colorlinks=true,linkcolor=blue,citecolor=blue,anchorcolor=blue,urlcolor=blue}

\makeatletter
\usepackage{dcolumn}
\usepackage{bm}
\usepackage{physics}
\usepackage{pifont}
\usepackage{fancyhdr}
\usepackage{color}
\usepackage{textcomp}

\usepackage[toc,page]{appendix}

\renewcommand{\fnum@figure}{FIG.~\thefigure}

\makeatother

\begin{document}
\title{Enantiodetection via the 2D spectroscopy: extending the methodology
to general experimental conditions}
\author{Mao-Rui Cai}
\affiliation{Graduate School of China Academy of Engineering Physics, No. 10 Xibeiwang
East Road, Haidian District, Beijing 100193, China}
\author{Chong Ye}
\affiliation{Beijing Key Laboratory of Nanophotonics and Ultrafine Optoelectronic
Systems School of Physics, Beijing Institute of Technology, Beijing
100081, China}
\author{Yong Li}
\email{yongli@hainanu.edu.cn}

\affiliation{Center for Theoretical Physics and School of Science, Hainan University,
Haikou 570228, China}
\author{Hui Dong}
\email{hdong@gscaep.ac.cn}

\affiliation{Graduate School of China Academy of Engineering Physics, No. 10 Xibeiwang
East Road, Haidian District, Beijing 100193, China}
\begin{abstract}
Developing effective methods to measure the enantiomeric excess of
the chiral mixture is one of the major topics in chiral molecular
researches, yet remains challenging. Enantiodetection method via two-dimensional
(2D) spectroscopy based on a four level model, containing a cyclic
three-level system (CTLS), of chiral molecules was recently proposed
and demonstrated, yet with a strict condition of the one-photon resonance
(where three driving fields are exactly resonantly coupled to the
three electric-dipole transitions, respectively) in the CTLS and narrowband
probe pulse assumption. Here, we extend the 2D spectroscopy method
to more general experimental conditions, with three-photon resonance
(where the sum of the two smaller frequencies among the three driving
fields equals to the third one) and broadband probe pulse. Our method
remains effective on enantiodetection with the help of experimental
techniques, such as the chop detection method, which is used to eliminate
the influence of the other redundant levels existing in the real system
of chiral molecules. Under these more general conditions, the enantiomeric
excess of the chiral mixture is estimated by taking an easily available
standard sample (usually the racemic mixture) as the reference.
\end{abstract}
\maketitle

\section{Introduction}

Chiral molecules have two species, i.e., enantiomers, which are mirror
images of each other. Enantiomers have almost the same physical properties,
e.g., energy level structures, yet act disparately in their biological
activities and chemical interaction~\citep{Quack1989}, leading to
successive interest on the investigation of enantioseparation~\citep{LiYong2007,LiXuan2010,Buhmann2019,LiuBo2021},
enantioconversion~\citep{Shapiro2000,Kral2003,YeChong2021JPB,YeChong2021PRA},
and enantiodetection~\citep{HeYanan2011,Stephens1985,Begzjav2019,Fanood2015,Ghosh2006,Patterson2013Nature,Patterson2013PRL,Lobsiger2015,Shubert2016,YeChong2019,Cai2022}.
Enantiomers often have subtle difference in their optical activities,
which is an exception to their identical physical properties~\citep{HeYanan2011,Stephens1985,Begzjav2019}.
This makes it possible to detect the enantiomeric excess of a chiral
mixture optically. However, traditional spectroscopic methods, such
as circular dichroism~\citep{HeYanan2011,Stephens1985} and Raman
optical activity~\citep{Begzjav2019}, typically suffer from weak
signal originated from magnetic-dipole or electric-quadrupole interactions.

In recent years, new spectroscopic methods of enantiodetection taking
advantage of the strong electric-dipole coupling have been investigated
theoretically~\citep{Jia2011,YeChong2019,YeChong2021JPCL,Cai2022}
and experimentally~\citep{Patterson2013Nature,Patterson2013PRL,Lobsiger2015,Shubert2016},
benefiting from the study of the cyclic three-level system (CTLS)
of the chiral molecules~\citep{Kral2001,Jacob2012,Lehmann2018,YeChong2018,Leibscher2019,ZhangQS2020},
where three electromagnetic driving fields are coupled to the three
electric-dipole transitions among three energy levels of the chiral
molecules. Typically, most of these methods require a standard enantiopure
sample as the reference~\citep{Patterson2013Nature,Patterson2013PRL,Lobsiger2015,Shubert2016}.
However, acquiring the enantiopure sample itself remains challenging
for most chiral molecules. We have recently proposed new enantiodetection
methods that do not require enantiopure sample, using either one-dimensional
(1D) or two-dimensional (2D) spectroscopy~\citep{Cai2022,YeChong2019}.
In the method via the 2D spectroscopy, three probe pulses are applied
in a sequence to detect the chirality-dependent energy shifts engineered
by the three electromagnetic driving fields in the CTLS~\citep{Cai2022,YeChong2019}.
Signals from different enantiomers are naturally categorized in the
2D spectrum as any two diagonal peaks corresponding to different enantiomers
have no off-diagonal correspondence~\citep{Jonas2003,Cai2022}. This
feature of the 2D spectroscopy provides advantage over the 1D case
, where additional procedures are required for peak categorization. 

We have demonstrated the effectiveness of our enantiodetection method~\citep{Cai2022}
via the 2D spectroscopy within ideal experimental conditions, e.g.,
the one-photon resonance~\citep{Vitanov2019} of the driving fields
in the CTLS and the use of narrowband probe pulses for single transition
excitation. However, in the case of one-photon resonance, all three
driving fields are resonantly coupled to the three electric-dipole
transitions without detuning. This demands a full understanding of
the energy levels of the investigated chiral molecules, limiting the
universal application of this method on enantiodetection. Moreover,
generating narrowband pulses covering only one transition is challenging.
In existing experiments of 2D spectroscopy, the bandwidth of the probe
pulses is usually broad, and multiple transitions from the ground
level to the excited levels can be induced~\citep{Reimann2021,Woerner2013,Lu2016},
which poses a challenge of this method on enantiodetection, as induced
multiple transitions may affect the ability of the 2D spectroscopy
to distinguish between different enantiomers.

In this paper we extend the 2D spectroscopy based enantiodetection
method to more general cases. Specifically, we demonstrate that different
enantiomers can still be distinguished by their separate and chirality-dependent
peaks in the 2D spectrum even under three-photon resonance~\citep{Kral2001,YeChong2019}
and with the use of broadband probe pulses. Under three-photon resonance,
the three driving fields are coupled to the corresponding transitions
with detunings, and the sum of the two smaller frequencies of the
driving fields equals to the third one. This condition is preferred
in the experiments, compared to the one-photon resonance, when one
has only rough knowledge of the interested enantiomers. We demonstrate
that under three-photon resonance, the enantiomeric excess of a chiral
mixture is estimated by taking the easy-to-get racemic mixture as
the reference. In more general situations where transitions from the
ground state to all three working excited states, as well as other
redundant states, are induced by broadband probe pulses, we prove
that these induced multiple transitions do not affect the enantiodetection
via the 2D spectrum. To achieve the enantiodetection in these general
situations, we adopt the advanced experimental techniques, i.e., the
chop detection~\citep{Reimann2021,Woerner2013,Lu2016} and truncation
method to eliminate the influence of the other redundant levels.

\section{Three-photon resonance\label{sec-TPR}}

\begin{figure}
\includegraphics[scale=0.52]{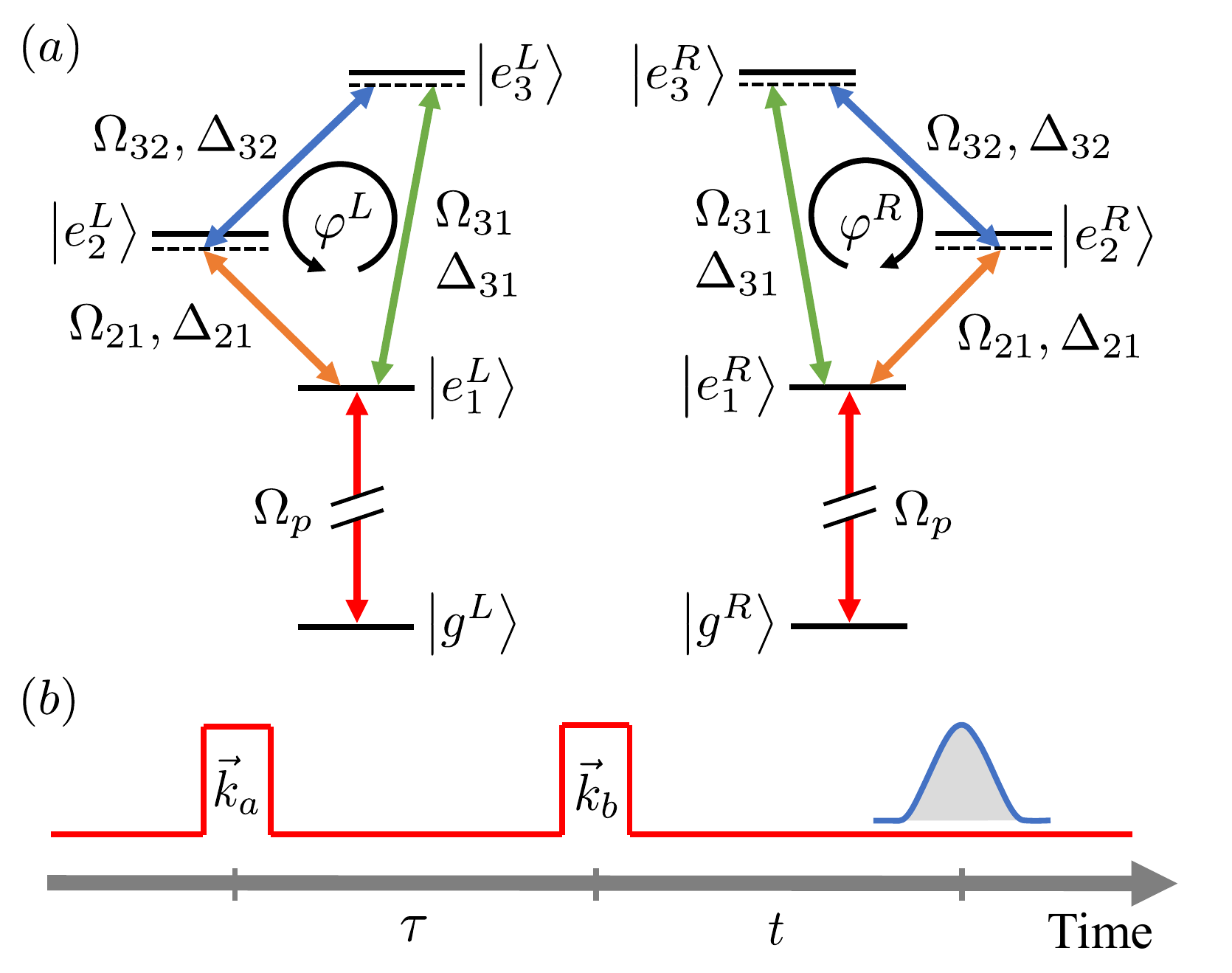}

\caption{(a) The basic model with a CTLS and a ground level of the chiral molecules.
In the subsystem, three driving fields are near resonantly coupled
to the three transitions $\ket{e_{j}^{\alpha}}\leftrightarrow\ket{e_{l}^{\alpha}}$
with Rabi frequencies $\Omega_{jl}$, detunings $\Delta_{jl}$, and
overall phases $\varphi^{\alpha}$. The probe pulses $\vec{k}_{p}$
($p=a,b$) induce only transition $\ket{g^{\alpha}}\leftrightarrow\ket{e_{1}^{\alpha}}$
with Rabi frequencies $\Omega_{p}$. (b) The probe pulse sequence.
Two probe pulse $\vec{k}_{a}$ and $\vec{k}_{b}$ denoted by the squares
are separated by an interval $\tau$. The signal denoted by the curve
is detected at time $t$ after the second probe pulse $\vec{k}_{b}$.\label{model}}
\end{figure}

The CTLS, in which three electric-dipole transitions among the three
energy levels are all allowed, is forbidden in the natural atoms,
but universally exists in the chiral molecules whose symmetry is naturally
broken. With this fact, we propose the basic model of this paper in
Fig.~\ref{model}(a). The model consists of a CTLS with three excited
states $\ket{e_{j}^{\alpha}}\:(j=1,2,3)$ and a ground state $\ket{g^{\alpha}}$
for left-handed ($\alpha=L$) and right-handed ($\alpha=R$) chiral
molecules~\citep{YeChong2021JPCL,ChenYuYuan2020,Eibenberger2017}.
Three constant electromagnetic driving fields (typically in the microwave
region) with central frequencies $\nu_{jl}$ ($j>l$) are applied
to couple the corresponding electric-dipole transitions $\ket{e_{j}^{\alpha}}\leftrightarrow\ket{e_{l}^{\alpha}}$
with detunings $\Delta_{jl}=\omega_{j}-\omega_{l}-\nu_{jl}$ under
three-photon resonance ($\nu_{21}+\nu_{32}=\nu_{31}$, i.e., $\Delta_{21}+\Delta_{32}=\Delta_{31}$).
Here, $\omega_{j}$ are eigenenergies of states $\ket{e_{j}}$ with
ground state energy $\omega_{0}=0$. The Hamiltonian of the system
is given in the interaction picture with respect to $H_{\alpha}'=\sum_{j}\omega_{j}'\ket{e_{j}^{\alpha}}\bra{e_{j}^{\alpha}}$
as ($\hbar=1$, $\omega_{1}'=\omega_{1}$, $\omega_{2}'=\omega_{1}+\nu_{21}$
and $\omega_{3}'=\omega_{1}+\nu_{31}$)

\begin{equation}
\begin{aligned}V_{\mathrm{cyc}}^{\alpha}= & \Delta_{21}\ket{e_{2}^{\alpha}}\bra{e_{2}^{\alpha}}+\Delta_{31}\ket{e_{3}^{\alpha}}\bra{e_{3}^{\alpha}}+(\Omega_{21}\ket{e_{2}^{\alpha}}\bra{e_{1}^{\alpha}}\\
+ & \Omega_{31}\ket{e_{3}^{\alpha}}\bra{e_{1}^{\alpha}}+\Omega_{32}e^{i\varphi^{\alpha}}\ket{e_{3}^{\alpha}}\bra{e_{2}^{\alpha}}+\mathrm{H.c.}).
\end{aligned}
\end{equation}
Here, $\Omega_{jl}$ are Rabi frequencies, and $\varphi^{\alpha}$
is the effective overall phase which differs by $\pi$ for the different
enantiomers~\citep{Kral2001,Jacob2012}, i.e., $\varphi^{R}=\varphi^{L}+\pi$.

In the proposed model, two pulses with wave vector $\vec{k}_{p}$
($p=a,b$), short duration $\delta t_{p}$, and interval $\tau$ are
applied to probe the chirality-dependent shifts induced by the driving
fields~\citep{YeChong2019}. The pulse sequence is illustrated in
Fig.~\ref{model}(b). The assumption is made that the probe pulses
induce only the transition $\ket{g^{\alpha}}\leftrightarrow\ket{e_{1}^{\alpha}}$
with central frequency $\nu_{p}=\omega_{1}$ and narrow bandwidth
$\delta\nu_{p}\ll\{\omega_{21},\omega_{31}\}$ where $\omega_{ij}=\left|\omega_{i}-\omega_{j}\right|$.
The Hamiltonian within the short duration of the probe pulse is given
in the interaction picture with respect to $H_{0}^{\alpha}=\sum_{j}\omega_{j}\ket{e_{j}^{\alpha}}\bra{e_{j}^{\alpha}}$
as
\begin{equation}
V_{p}^{\alpha}=\Omega_{p}e^{i\vec{k}_{p}\cdot\vec{r}}\ket{e_{1}^{\alpha}}\bra{g^{\alpha}}+\mathrm{H.c.},
\end{equation}
where $\vec{r}$ is the spatial location of the chiral molecule and
$\Omega_{p}$ is the Rabi frequency corresponding to transition $\ket{g^{\alpha}}\leftrightarrow\ket{e_{1}^{\alpha}}$
under the square pulse approximation. We have neglected the interaction
of the CTLS during the short duration of the probe pulses because
the driving fields are much weaker than the probe pulses, i.e., $\left|\Omega_{jl}\right|\ll\left|\Omega_{p}\right|$.

In our previous work in Ref.~\citep{Cai2022}, a strict condition
of one-photon resonance ($\Delta_{jl}=0$) and narrow bandwidth probe
pulse assumption were considered. Here, we release only the condition
of one-photon resonance to show how the properties of driving field
would impact the shape of the 2D spectra. We will later show in Sec.~\ref{BroadBW}
that pulses with broad bandwidth could also be applied as probe pulses.

Starting from the initial state $\ket{\psi_{0}^{\alpha}}=\ket{g^{\alpha}}$
in the Schr\"{o}dinger picture, the final state of the system after
interaction with the sequence of the two pulses is
\begin{equation}
\ket{\psi^{\alpha}(\tau,t)}=U_{\mathrm{cyc}}^{\alpha}(t)U_{b}^{\alpha}U_{\mathrm{cyc}}^{\alpha}(\tau)U_{a}^{\alpha}\ket{\psi_{0}^{\alpha}},
\end{equation}
where $U_{p}^{\alpha}=\exp[-iH_{0}^{\alpha}\delta t_{p}]\exp[-iV_{p}^{\alpha}\delta t_{p}]$
and $U_{\mathrm{cyc}}^{\alpha}(s)=\exp[-iH_{\alpha}'s]\exp[-iV_{\mathrm{cyc}}^{\alpha}s]$
are evolution operators inside and outside the pulse duration, respectively.
The final polarization $\bm{P}^{\alpha}(\tau,t)\equiv\bra{\psi^{\alpha}(\tau,t)}\hat{\bm{\mu}}^{\alpha}\ket{\psi^{\alpha}(\tau,t)}$
will yield signal fields along multiple phase-matching directions~\citep{mukamel1995,Cho2009,Tian2003},
and we sort only the rephasing signal that emits along $-\vec{k}_{a}+2\vec{k}_{b}$. 

In the final polarization, the rephasing part is calculated as $\bm{P}_{\mathrm{RP}}^{\alpha}(\tau,t)=\bm{A}^{\alpha}(\tau,t)e^{i(-\vec{k}_{a}+2\vec{k}_{b})\cdot\vec{r}}$.
Here,
\begin{equation}
\begin{aligned}\bm{A}^{\alpha}(\tau,t)= & \mathcal{N}_{a}^{2}\mathcal{N}_{b}^{2}\beta_{a}^{*}\left|\beta_{b}\right|^{2}\sum_{j'}\left|n_{1j'}^{\alpha}\right|^{2}e^{i(\omega_{1}'+E_{j'}^{\alpha})\tau}\\
\times & \sum_{j}\left|n_{1j}^{\alpha}\right|^{2}e^{-i(\omega_{1}'+E_{j}^{\alpha})t}\bm{\mu}_{01},
\end{aligned}
\label{sig-rp0}
\end{equation}
is the amplitude of the rephasing signal, $\hat{\bm{\mu}}^{\alpha}$
is transition dipole operator, $\bm{\mu}_{01}^{\alpha}=\bm{\mu}_{01}$
is the transition dipole moment corresponding to the transition $\ket{g^{\alpha}}\leftrightarrow\ket{e_{1}^{\alpha}}$,
$\beta_{p}=-i\Omega_{p}\delta t_{p}$ are transition amplitudes, $\mathcal{N}_{p}=(1+\left|\beta_{p}\right|^{2})^{-1/2}$
are normalized constants, $n_{lj}=\braket{e_{l}^{\alpha}}{d_{j}^{\alpha}}$
are transformation matrix elements, and $\ket{d_{j}^{\alpha}}$ and
$E_{j}^{\alpha}$ are eigenvectors and eigenvalues of $V_{\mathrm{cyc}}^{\alpha}$.
Adding the decoherence of the system during the whole process, the
rephasing signal in Eq.~(\ref{sig-rp0}) is modified as
\begin{equation}
\begin{aligned}\bm{A}^{\alpha}(\tau,t)= & \mathcal{N}_{a}^{2}\mathcal{N}_{b}^{2}\beta_{a}^{*}\left|\beta_{b}\right|^{2}\sum_{j'}\left|n_{1j'}^{\alpha}\right|^{2}e^{i(\omega_{1}'+E_{j'}^{\alpha})\tau}e^{-\Gamma\tau}\\
\times & \sum_{j}\left|n_{1j}^{\alpha}\right|^{2}e^{-i(\omega_{1}'+E_{j}^{\alpha})t}e^{-\Gamma t}\bm{\mu}_{01},
\end{aligned}
\label{sig-rp1}
\end{equation}
where $\Gamma=\gamma_{\mathrm{rl}}/2+\gamma_{\mathrm{dp}}$ is the
decoherence rate, $\gamma_{\mathrm{rl}}$ and $\gamma_{\mathrm{dp}}$
are correspondingly the relaxation and the pure dephasing rates of
the excited states.

For a mixture, the total signal from the two enantiomers is
\begin{equation}
\bm{A}^{\mathrm{mix}}(\tau,t)=N_{L}\bm{A}^{L}(\tau,t)+N_{R}\bm{A}^{R}(\tau,t),
\end{equation}
where $N_{L}$ and $N_{R}$ are numbers of the left- and right-handed
molecules in the mixture sample. The 2D spectrum
\begin{equation}
\tilde{\bm{A}}^{\mathrm{mix}}(\omega_{\tau},\omega_{t})\equiv\mathcal{F}\left[\bm{A}^{\mathrm{mix}}(\tau,t)\right],
\end{equation}
is obtained by Fourier transforming~\citep{Fleming-CP2011} $\bm{A}^{\mathrm{mix}}(\tau,t)$
with respect to both the delay time $\tau$ and data collecting time
$t$.

\begin{figure}[tbh]
\includegraphics{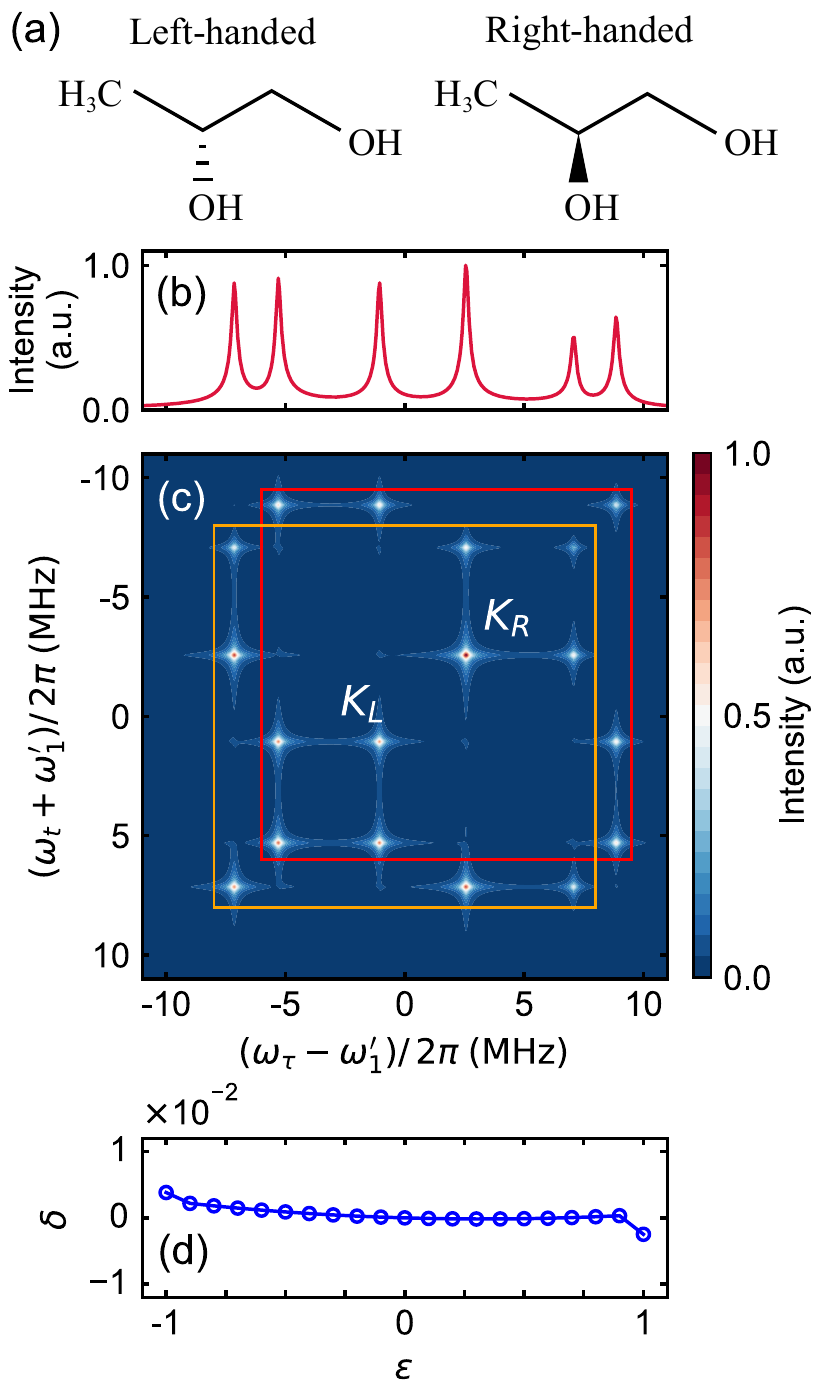}

\caption{(a) The skeletal formulas of two enantiomers of 1,2-propanediol which
is taken as the example. (b) 1D and (c) 2D spectra of the racemic
mixture (50:50 mixture of two enantiomers) of gaseous 1,2-propanediol.
The 1D spectrum is the projection of the 2D spectrum on the axis of
$\omega_{\tau}$. The 2D spectrum is obtained by taking only the absolute
value of the Fourier transform. $K_{L}$ and $K_{R}$ in (c) denote
the amplitudes of two strongest peaks that belong to left-handed and
right-handed enantiomers, respectively. (d) The error $\delta=\varepsilon_{e}-\varepsilon$
between the enantiomeric excess $\varepsilon_{e}$ estimated by the
intensity ratio of the peaks denoted in (c) and the real value $\varepsilon$.
The parameters for the simulation are chosen as $\Omega_{p}/2\pi\simeq100\:\mathrm{MHz}$,
$\Omega_{21}/2\pi=4\:\mathrm{MHz}$, $\Omega_{31}/2\pi=4\:\mathrm{MHz}$,
$\Omega_{32}/2\pi=4.5\:\mathrm{MHz}$, $\Delta_{21}/2\pi=1\:\mathrm{MHz}$,
$\Delta_{31}/2\pi=1.5\:\mathrm{MHz}$, $\gamma_{\mathrm{rl}}/2\pi\simeq1\:\mathrm{kHz}$,
$\gamma_{\mathrm{dp}}/2\pi\simeq0.1\:\mathrm{MHz}$, $\varphi^{L}=2\pi/7$
and $\varphi^{R}=9\pi/7$. \label{2D-3-photon-resn}}
\end{figure}

We take the gaseous 1,2-propanediol whose skeletal formula is shown
in Fig.~\ref{2D-3-photon-resn}(a) as our example in this study and
present a numerically simulated 2D spectrum of the racemic mixture
(50:50 mixture of two enantiomers) in Fig.~\ref{2D-3-photon-resn}(c)
which is zoomed around $(\omega_{1}',-\omega_{1}')$. Neglecting the
chirality index, the working states are chosen as $\ket{g}=\ket{v_{g}}\ket{0_{0,0,0}}$,
$\ket{e_{1}}=\ket{v_{e}}\ket{1_{1,1,1}}$, $\ket{e_{2}}=\ket{v_{e}}\ket{2_{2,1,2}}$,
and $\ket{e_{3}}=\ket{v_{e}}\ket{2_{2,0,1}}$, where the vibrational
ground (first-excited) state is denoted as $\ket{v_{g}}$ ($\ket{v_{e}}$)
and the rotational states are denoted in the $\ket{J_{K_{a},K_{c},M}}$
notation~\citep{Zare1988}. With these working states, the transition
frequencies are $\omega_{10}/2\pi\simeq4.33\:\mathrm{THz}$, $\omega_{21}/2\pi\simeq29.21\:\mathrm{GHz}$,
$\omega_{31}/2\pi\simeq29.31\:\mathrm{GHz}$, and $\omega_{32}/2\pi\simeq100.76\:\mathrm{MHz}$~\citep{LOVAS2009,Cai2022,ARENAS2017}.
The relaxation and pure dephasing rates are approximately taken as
$\gamma_{\mathrm{rl}}/2\pi\simeq1\:\mathrm{kHz}$ and $\gamma_{\mathrm{dp}}/2\pi\simeq0.1\:\mathrm{MHz}$~\citep{Eibenberger2017,Patterson2012}.
The CTLS are assumed to have detunings $\Delta_{21}/2\pi=1\:\mathrm{MHz}$
and $\Delta_{31}/2\pi=1.5\:\mathrm{MHz}$ of the driving fields, Rabi
frequencies $\Omega_{21}/2\pi=4\:\mathrm{MHz}$, $\Omega_{31}/2\pi=4\:\mathrm{MHz}$,
and $\Omega_{32}/2\pi=4.5\:\mathrm{MHz}$, and overall phases $\varphi^{L}=2\pi/7$
and $\varphi^{R}=9\pi/7$. The probe pulses have duration $\delta t_{p}\simeq0.5\:\mathrm{ns}$
with Fourier limited bandwidth $\delta\nu\simeq2\pi\times1.8\:\mathrm{GHz}\ll\left\{ \omega_{21},\omega_{31}\right\} $
and corresponding Rabi frequencies $\Omega_{p}/2\pi\simeq100\:\mathrm{MHz}$.
The time domain signal is scanned from 0 to $20\:\mu\mathrm{s}$ with
step size $0.01\:\mu\mathrm{s}$.

According to Eq.~(\ref{sig-rp1}), each enantiomer has nine peaks
in the 2D spectrum in Fig.~\ref{2D-3-photon-resn}(c) at locations
$(\omega_{1}'+E_{j'}^{\alpha},-\omega_{1}'-E_{j}^{\alpha})$. Those
nine peaks form a grouped pattern, with three diagonal peaks resulting
in six off-diagonal ones. The lack of off-diagonal correspondence
between any two diagonal peaks indicates that they belong to opposite
chiralities. For instance, nine peaks enclosed by red lines in Fig.~\ref{2D-3-photon-resn}(c)
share the same chirality (i.e., left-handed chirality), and the other
nine peaks enclosed by orange lines share the opposite (i.e., right-hand
chirality). We note that such a direct categorization is not possible
within a single 1D spectrum, e.g., the one in Fig.~\ref{2D-3-photon-resn}(b),
without additional procedure~\citep{YeChong2019,Jia2011}.

We remark that the distinct nine peaks pattern is the general case,
and different patterns less than nine peaks may appear due to the
degeneracy of $E_{j}^{\alpha}$~\citep{Cai2022} or the inequality
of $\left|n_{1j}^{\alpha}\right|^{2}$. Nevertheless, the chirality
categorization can always be achieved by pointing the off-diagonal
peaks.

After the chirality categorization, the enantiomeric excess $\varepsilon=(N_{L}-N_{R})/(N_{L}+N_{R})$
of the chiral mixture is estimated by the intensity ratio of the peaks
in the 2D spectrum. To clarify the notation used in our analysis,
we denote the amplitudes of the two strongest diagonal peaks belonging
to different enantiomers at locations $(\omega_{L},-\omega_{L})$
and $(\omega_{R},-\omega_{R})$ as $K_{L}$ and $K_{R}$, where
\begin{equation}
K_{\alpha}=N_{L}\left|\tilde{\bm{A}}^{L}(\omega_{\alpha},-\omega_{\alpha})\right|+N_{R}\left|\tilde{\bm{A}}^{R}(\omega_{\alpha},-\omega_{\alpha})\right|,
\end{equation}
$\omega_{\alpha}=\omega_{1}'+E_{2}^{\alpha}$ with $E_{2}^{L}/2\pi=-1.060\:\mathrm{MHz}$
and $E_{2}^{R}/2\pi=2.566\:\mathrm{MHz}$. Since the peaks belonging
to different enantiomers are well separated, left-handed enantiomer
has very little contributions to $K_{R}$ and vice versa, i.e., $\left|\tilde{\bm{A}}^{L}(\omega_{L},-\omega_{L})\right|\gg\left|\tilde{\bm{A}}^{R}(\omega_{L},-\omega_{L})\right|$,
$\left|\tilde{\bm{A}}^{R}(\omega_{R},-\omega_{R})\right|\gg\left|\tilde{\bm{A}}^{L}(\omega_{R},-\omega_{R})\right|$,
and
\begin{equation}
K_{\alpha}\simeq N_{\alpha}\left|\tilde{\bm{A}}^{\alpha}(\omega_{\alpha},-\omega_{\alpha})\right|.
\end{equation}

In the one-photon resonance case ($\Delta_{jl}=0$) of the CTLS, the
enantiopure signal intensities are always equal $\left|\tilde{\bm{A}}^{L}(\omega_{L},-\omega_{L})\right|=\left|\tilde{\bm{A}}^{R}(\omega_{R},-\omega_{R})\right|$
due to the equality of the absolute value~\citep{Cai2022} of the
corresponding transformation matrix elements $\left|n_{1j}^{L}\right|=\left|n_{1j}^{R}\right|$.
The enantiomeric excess is estimated using the equation $\varepsilon_{e}=(K_{L}-K_{R})/(K_{L}+K_{R})$.

However, in the general three-photon resonance case ($\Delta_{21}+\Delta_{32}=\Delta_{31}$),
the absolute values of the transformation matrix elements are typically
not equal and are often unknown beforehand. Therefore, we use the
racemic mixture of chiral molecules as the reference sample, which
is easily accessible and readily available, to obtain a parameter
\begin{equation}
\lambda=\frac{K_{L}^{\mathrm{rm}}}{K_{R}^{\mathrm{rm}}},
\end{equation}
and the enantiomeric excess is estimated as
\begin{equation}
\varepsilon_{e}=\frac{K_{L}-\lambda K_{R}}{K_{L}+\lambda K_{R}}.\label{est-ee}
\end{equation}
We demonstrate the numerical estimation error $\delta=\varepsilon_{e}-\varepsilon$
as a function of $\varepsilon$ in Fig.~\ref{2D-3-photon-resn}(d)
and we observe that the absolute value of the error is consistently
less than $0.5\times10^{-2}$.

Since the one-photon resonance case requires a full understanding
of the transition frequencies in the CTLS, which is a strict condition
in experiments, we prove in this section that our method is capable
of enantiodetection under the general three-photon resonance using
the racemic mixture sample as the reference to estimate the enantiomeric
excess.

\section{Broadband probe pulse\label{BroadBW}}

In the above discussion, we have assumed that only one transition
$\ket{g^{\alpha}}\leftrightarrow\ket{e_{1}^{\alpha}}$ is induced
by the probe pulses with the narrowband assumption. However, in experimental
conditions, the pulse bandwidth is usually much broader, on the order
of several terahertz~\citep{Reimann2021,Woerner2013,Lu2016}, which
means that all three transitions $\ket{g^{\alpha}}\leftrightarrow\ket{e_{j}^{\alpha}}$
may be induced (if the selection rule ~\citep{YeChong2018,Jacob2012,Leibscher2019}
also permits). Despite the specifically chosen working states, some
other redundant states can get involved in the probing process as
noise.

We prove in this section that the chiral molecules are still distinguishable
in the 2D spectrum even if more than one transition are induced by
the probe pulses. Furthermore, the influence of other redundant states
is eliminated by the chop detection method, which is typically used
in optical experiments~\citep{Reimann2021,Woerner2013,Lu2016}.

\subsection{Inducing transitions to multiple excited working states\label{subsec-ThreePumping}}

Instead of narrowband assumption, we here and hereafter assume that
the bandwidth of the probe pulses is so broad that all transitions
$\ket{g^{\alpha}}\leftrightarrow\ket{e_{j}^{\alpha}}$ are nearly
resonantly induced~\citep{ZhangXue2022}. Thus, the Hamiltonian of
the system within the pulse duration is given in the interaction picture
with respect to $H_{0}^{\alpha}$ as
\begin{equation}
V_{p,\mathrm{3e}}^{\alpha}=\sum_{j}\Omega_{p,j}^{\alpha}e^{i\vec{k}_{p}\cdot\vec{r}}\ket{e_{j}^{\alpha}}\bra{g^{\alpha}}+\mathrm{H.c.},
\end{equation}
and the rephasing signal is correspondingly modified as
\begin{equation}
\begin{aligned}\bm{A}_{\mathrm{3e}}^{\alpha}(\tau,t)= & \left[\mathcal{N}_{a,0}^{\alpha}\right]^{2}\mathcal{N}_{b,0}^{\alpha}\sum_{m,m'}\left[\beta_{a,m}^{\alpha}\right]^{*}\beta_{b,m'}^{\alpha}\\
\times & \sum_{j,l}[n_{lj}^{\alpha}]^{*}n_{mj}^{\alpha}e^{i(\omega_{l}'+E_{j}^{\alpha})\tau}e^{-\Gamma\tau}\mathcal{N}_{b,l}^{\alpha}\left[\beta_{b,l}^{\alpha}\right]^{*}\\
\times & \sum_{j',l'}n_{l'j'}^{\alpha}[n_{m'j'}^{\alpha}]^{*}e^{-i(\omega_{l'}'+E_{j'}^{\alpha})t}e^{-\Gamma t}\bm{\mu}_{0l'}^{\alpha}.
\end{aligned}
\label{sig-rp-3pump}
\end{equation}
Here, $\bm{\mu}_{0j}^{\alpha}$ and $\Omega_{p,j}^{\alpha}$ are transition
dipole moments and Rabi frequencies corresponding to transitions $\ket{g^{\alpha}}\leftrightarrow\ket{e_{j}^{\alpha}}$,
$\beta_{p,j}^{\alpha}=-i\Omega_{p,j}^{\alpha}\delta t_{p}$ are transition
amplitudes, $\mathcal{N}_{p,j}^{\alpha}=(1+\left|\beta_{p,j}^{\alpha}\right|^{2})^{-1/2}$
and $\mathcal{N}_{p,0}^{\alpha}=(1+\sum_{j}\left|\beta_{p,j}^{\alpha}\right|^{2})^{-1/2}$
are normalized constants. We specify the transition dipole moments
as $\bm{\mu}_{01}^{L}=\bm{\mu}_{01}^{R}=\bm{\mu}_{01}$, $\bm{\mu}_{02}^{L}=-\bm{\mu}_{02}^{R}=\bm{\mu}_{02}$
and $\bm{\mu}_{03}^{L}=-\bm{\mu}_{03}^{R}=\bm{\mu}_{03}$ to fulfill
the cyclic conditions of the new cyclic structures formed by levels
$\ket{g^{\alpha}}$, $\ket{e_{j}^{\alpha}}$ and $\ket{e_{j'}^{\alpha}}$
($j\neq j'$). For the special case with the selection rule considered,
one or two of the transitions $\ket{g^{\alpha}}\leftrightarrow\ket{e_{j}^{\alpha}}$
may be forbidden, i.e., $\bm{\mu}_{0j}=0$. Especially, when two of
the three transitions are forbidden, the rephasing signal in Eq.~(\ref{sig-rp-3pump})
retains the one in Eq.~(\ref{sig-rp1}) where only one transition
is considered.

The 2D spectrum is obtained by Fourier transforming $\bm{A}_{\mathrm{3e}}^{\alpha}(\tau,t)$
with respect to $\tau$ and $t$. And the peaks still locate at positions
$(\omega_{1}'+E_{j'}^{\alpha},-\omega_{1}'-E_{j}^{\alpha})$ because
the phase factors containing $\tau$ and $t$ in Eq.~(\ref{sig-rp-3pump})
are exactly the same with those in Eq.~(\ref{sig-rp1}). However,
the peak amplitudes change due to the different interaction.

\begin{figure}
\includegraphics{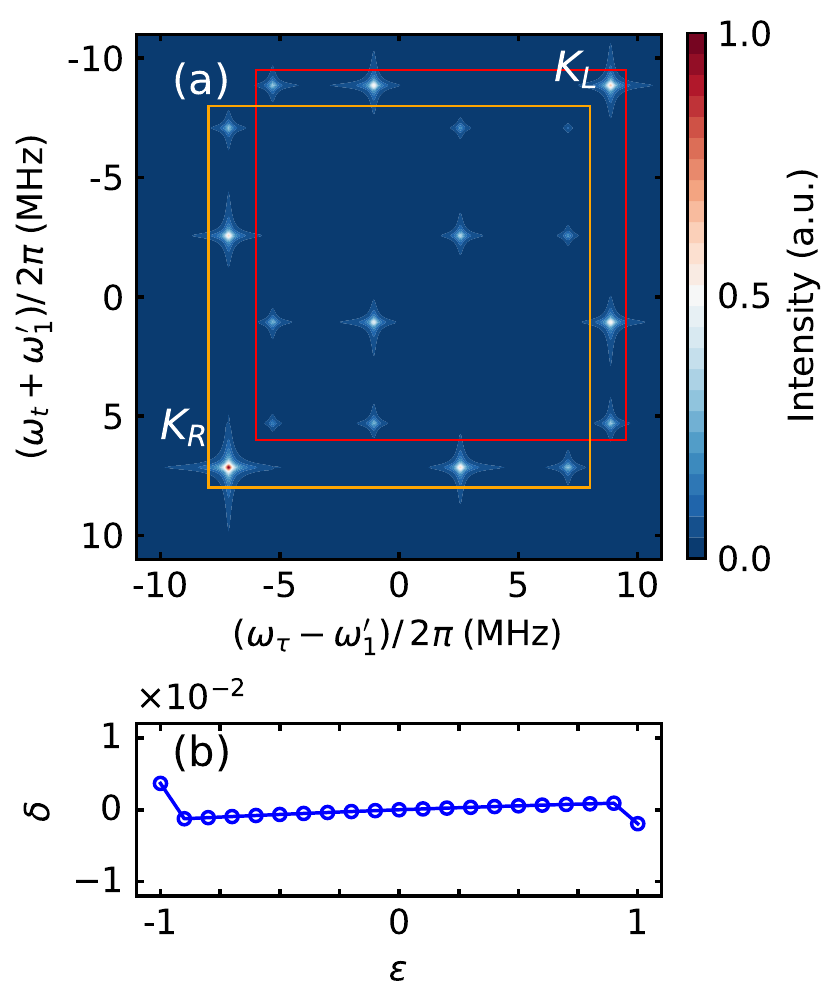}

\caption{(a) 2D spectrum of the racemic mixture of gaseous 1,2-propanediol
when both transition $\ket{g^{\alpha}}\leftrightarrow\ket{e_{1}^{\alpha}}$
and $\ket{g^{\alpha}}\leftrightarrow\ket{e_{2}^{\alpha}}$ are induced
by the probe pulses. (b) The error of the estimation of the enantiomeric
excess estimated by the intensity ratio of the peaks denoted in (a).
In the simulation, the Rabi frequencies between the ground state and
the excited states are chosen as $\Omega_{p,1}^{L}=\Omega_{p,1}^{R}=2\pi\times100\:\mathrm{MHz}$
and $\Omega_{p,2}^{L}=-\Omega_{p,2}^{R}=2\pi\times100\:\mathrm{MHz}$,
and the transition $\ket{g^{\alpha}}\leftrightarrow\ket{e_{3}^{\alpha}}$
is forbidden. The other parameters are taken the same with those in
Fig.~\ref{2D-3-photon-resn}. \label{2D-ThreePumping}}

\end{figure}

The simulated 2D spectrum of the racemic mixture with two transitions
$\ket{g^{\alpha}}\leftrightarrow\ket{e_{1}^{\alpha}}$ and $\ket{g^{\alpha}}\leftrightarrow\ket{e_{2}^{\alpha}}$
induced is presented in Fig.~\ref{2D-ThreePumping}(a) by taking
the working states of 1,2-propanediol as $\ket{g}=\ket{v_{g}}\ket{0_{0,0,0}}$,
$\ket{e_{1}}=\ket{v_{e}}\ket{1_{1,1,1}}$, $\ket{e_{2}}=\ket{v_{e}}\ket{1_{1,0,1}}$,
and $\ket{e_{3}}=\ket{v_{e}}\ket{2_{2,0,1}}$.

The transition frequencies in the CTLS are $\omega_{21}/2\pi\simeq846.79\:\mathrm{MHz}$,
$\omega_{31}/2\pi\simeq29.31\:\mathrm{GHz}$, and $\omega_{32}/2\pi\simeq28.46\:\mathrm{GHz}$~\citep{ARENAS2017,LOVAS2009}.
We maintain the same assumptions regarding the detunings, Rabi frequencies,
and overall phases of the CTLS as described in Sec.~\ref{sec-TPR}.
However, we modify the assumption of the probe pulses to have a bandwidth
of $\delta\nu/2\pi\simeq0.9\:\mathrm{THz}$ and central frequency
of $\nu_{p}/2\pi\simeq4.33\:\mathrm{THz}$. With such a broad bandwidth,
all three transitions $\ket{g^{\alpha}}\leftrightarrow\ket{e_{j}^{\alpha}}$
are covered, but the transition $\ket{g^{\alpha}}\leftrightarrow\ket{e_{3}^{\alpha}}$
is forbidden since $\Delta J=2$~\citep{YeChong2018,Leibscher2019,Jacob2012}.
Moreover, we take $\Omega_{p,1}^{L}=\Omega_{p,1}^{R}=2\pi\times100\:\mathrm{MHz}$
and $\Omega_{p,2}^{L}=-\Omega_{p,2}^{R}=2\pi\times100\:\mathrm{MHz}$.

We also denote the amplitudes of the two strongest diagonal peaks
belonging to different enantiomers as $K_{L}$ and $K_{R}$ in Fig.~\ref{2D-ThreePumping}(a).
Such two peaks locate at $\left(\omega_{L}',-\omega_{L}'\right)$
and $\left(\omega_{R}',-\omega_{R}'\right)$ respectively, where $\omega_{L}'=\omega_{1}'+E_{3}^{L}$
and $\omega_{R}'=\omega_{1}'+E_{1}^{R}$ with $E_{3}^{L}/2\pi=8.861\:\mathrm{MHz}$
and $E_{1}^{R}/2\pi=-7.145\:\mathrm{MHz}$. The estimation of enantiomeric
excess is obtained using Eq.~(\ref{est-ee}) by taking the racemic
mixture as the reference. Fig.~\ref{2D-ThreePumping}(b) shows that
the errors of such estimation are still well below $1\times10^{-2}$
even with broadband probe pulses.

\subsection{Inducing transitions to redundant excited states}

Apart from the chosen working states, real chiral molecules have complex
rotational levels~\citep{Lu2016}, which will yield massive redundant
peaks on the 2D spectrum, disturbing the categorization of the chirality
of the peaks. For the case in Sec.~\ref{subsec-ThreePumping}, transition
between the ground state and a redundant excited state $\ket{e_{\mathrm{rd}}}=\ket{v_{e}}\ket{1_{0,1,1}}$
could also be induced by the probe pulses. Such a state will not introduce
additional cyclic loop since the transition frequencies between $\ket{e_{\mathrm{rd}}}$
and $\ket{e_{j}^{\alpha}}$ ($2\pi\times4.89\:\mathrm{GHz}$, $2\pi\times5.74\:\mathrm{GHz}$
and $2\pi\times34.20\:\mathrm{GHz}$) are all significantly non-resonant
with the frequencies of the driving fields. We take into account this
redundant state and modify the Hamiltonian within the pulse duration
in the interaction picture with respect to $H_{0}^{\alpha}$ as $V_{p,\mathrm{me}}^{\alpha}=V_{p,\mathrm{3e}}^{\alpha}+V_{\mathrm{rd}}^{\alpha}$
where
\begin{equation}
V_{\mathrm{rd}}^{\alpha}=\Omega_{p,\mathrm{rd}}e^{i\vec{k}_{p}\cdot\vec{r}}\ket{e_{\mathrm{rd}}^{\alpha}}\bra{g^{\alpha}}+\mathrm{H.c.},
\end{equation}
and $\Omega_{p,\mathrm{rd}}$ is the Rabi frequency corresponding
to transition $\ket{g^{\alpha}}\leftrightarrow\ket{e_{\mathrm{rd}}^{\alpha}}$.
Given this Hamiltonian, the final rephasing signal is $\bm{A}_{\mathrm{me}}^{\alpha}=\bm{A}_{\mathrm{3e}}^{\alpha}+\bm{A}_{\mathrm{rd1}}^{\alpha}+\bm{A}_{\mathrm{rd2}}^{\alpha}$
with 
\begin{equation}
\begin{aligned}\bm{A}_{\mathrm{rd1}}^{\alpha}(\tau,t)= & \left[\mathcal{N}_{a,0}^{\alpha}\right]^{2}\mathcal{N}_{b,0}^{\alpha}\mathcal{N}_{b,\mathrm{rd}}\beta_{a,\mathrm{rd}}^{*}\left|\beta_{b,\mathrm{rd}}\right|^{2}e^{i\omega_{\mathrm{rd}}\tau}e^{-\Gamma\tau}\\
\times & e^{-i\omega_{\mathrm{rd}}t}e^{-\Gamma t}\bm{\mu}_{\mathrm{rd}},
\end{aligned}
\label{diag-rd}
\end{equation}
and
\begin{equation}
\begin{aligned}\bm{A}_{\mathrm{rd2}}^{\alpha}(\tau,t)= & \left[\mathcal{N}_{a,0}^{\alpha}\right]^{2}\mathcal{N}_{b,0}^{\alpha}\mathcal{N}_{b,\mathrm{rd}}\beta_{a,\mathrm{rd}}\beta_{b,\mathrm{rd}}\sum_{m}\beta_{b,m}^{\alpha}\\
\times & e^{i\omega_{\mathrm{rd}}\tau}e^{-\Gamma\tau}\sum_{j,l}n_{lj}^{\alpha}[n_{mj}^{\alpha}]^{*}e^{-i(\omega_{l}'+E_{j}^{\alpha})t}e^{-\Gamma t}\bm{\mu}_{0l}^{\alpha}\\
+ & \left[\mathcal{N}_{a,0}^{\alpha}\right]^{2}\mathcal{N}_{b,0}^{\alpha}\beta_{b,\mathrm{rd}}\sum_{m}\beta_{a,m}^{\alpha}\sum_{j,l}\mathcal{N}_{b,l}^{\alpha}\left[\beta_{b,l}^{\alpha}\right]^{*}\\
\times & [n_{lj}^{\alpha}]^{*}n_{mj}^{\alpha}e^{i(\omega_{l}'+E_{j}^{\alpha})\tau}e^{-\Gamma\tau}e^{-i\omega_{\mathrm{rd}}t}e^{-\Gamma t}\bm{\mu}_{\mathrm{rd}},
\end{aligned}
\end{equation}
where $\omega_{\mathrm{rd}}$ is the eigenenergy of the redundant
level $\ket{e_{\mathrm{rd}}^{\alpha}}$, $\beta_{p,\mathrm{rd}}=-i\Omega_{p,\mathrm{rd}}\delta t_{p}$
is the transition amplitudes, $\mathcal{N}_{p,j}^{\alpha}=(1+\left|\beta_{p,j}^{\alpha}\right|^{2})^{-1/2}$,
$\mathcal{N}_{p,\mathrm{rd}}=(1+\left|\beta_{p,\mathrm{rd}}\right|^{2})^{-1/2}$,
and $\mathcal{N}_{p,0}^{\alpha}=(1+\left|\beta_{p,\mathrm{rd}}\right|^{2}+\sum_{j}\left|\beta_{p,j}^{\alpha}\right|^{2})^{-1/2}$
are the normalized constants, and $\bm{\mu}_{\mathrm{rd}}^{\alpha}=\bm{\mu}_{\mathrm{rd}}$
is the transition dipole moment corresponding to transition $\ket{g^{\alpha}}\leftrightarrow\ket{e_{\mathrm{rd}}^{\alpha}}$.

According to $\bm{A}_{\mathrm{rd1}}^{\alpha}(\tau,t)$ in Eq.~(\ref{diag-rd}),
there is an additional diagonal peak at location $(\omega_{\mathrm{rd}},-\omega_{\mathrm{rd}})$
in the Fourier transformed 2D spectrum. This peak has no chiral feature
($\mathcal{N}_{p,0}^{L}=\mathcal{N}_{p,0}^{R}$) and is not affected
by the driving fields in the CTLS. We thus eliminate it through the
chop detection~\citep{Reimann2021,Woerner2013,Lu2016} by conducting
two experiments with driving fields on and off. The chop detected
signal is given in the frequency domain as 
\begin{equation}
\begin{aligned}\tilde{\bm{A}}_{\mathrm{chop}}^{\alpha}(\omega_{\tau},\omega_{t})= & \left|\tilde{\bm{A}}_{\mathrm{3e,\mathrm{on}}}^{\alpha}(\omega_{\tau},\omega_{t})\right|-\left|\tilde{\bm{A}}_{\mathrm{3e,\mathrm{off}}}^{\alpha}(\omega_{\tau},\omega_{t})\right|\\
+ & \left|\tilde{\bm{A}}_{\mathrm{\mathrm{rd2},\mathrm{on}}}^{\alpha}(\omega_{\tau},\omega_{t})\right|-\left|\tilde{\bm{A}}_{\mathrm{\mathrm{rd2},\mathrm{off}}}^{\alpha}(\omega_{\tau},\omega_{t})\right|.
\end{aligned}
\end{equation}
The two terms $\left|\tilde{\bm{A}}_{\mathrm{\mathrm{rd2},\mathrm{on}}}^{\alpha}(\omega_{\tau},\omega_{t})\right|$
and $\left|\tilde{\bm{A}}_{\mathrm{\mathrm{rd2},\mathrm{off}}}^{\alpha}(\omega_{\tau},\omega_{t})\right|$
have only off-diagonal peaks and thus would not disturb the chirality
categorization of the diagonal peaks.

\begin{figure}
\includegraphics{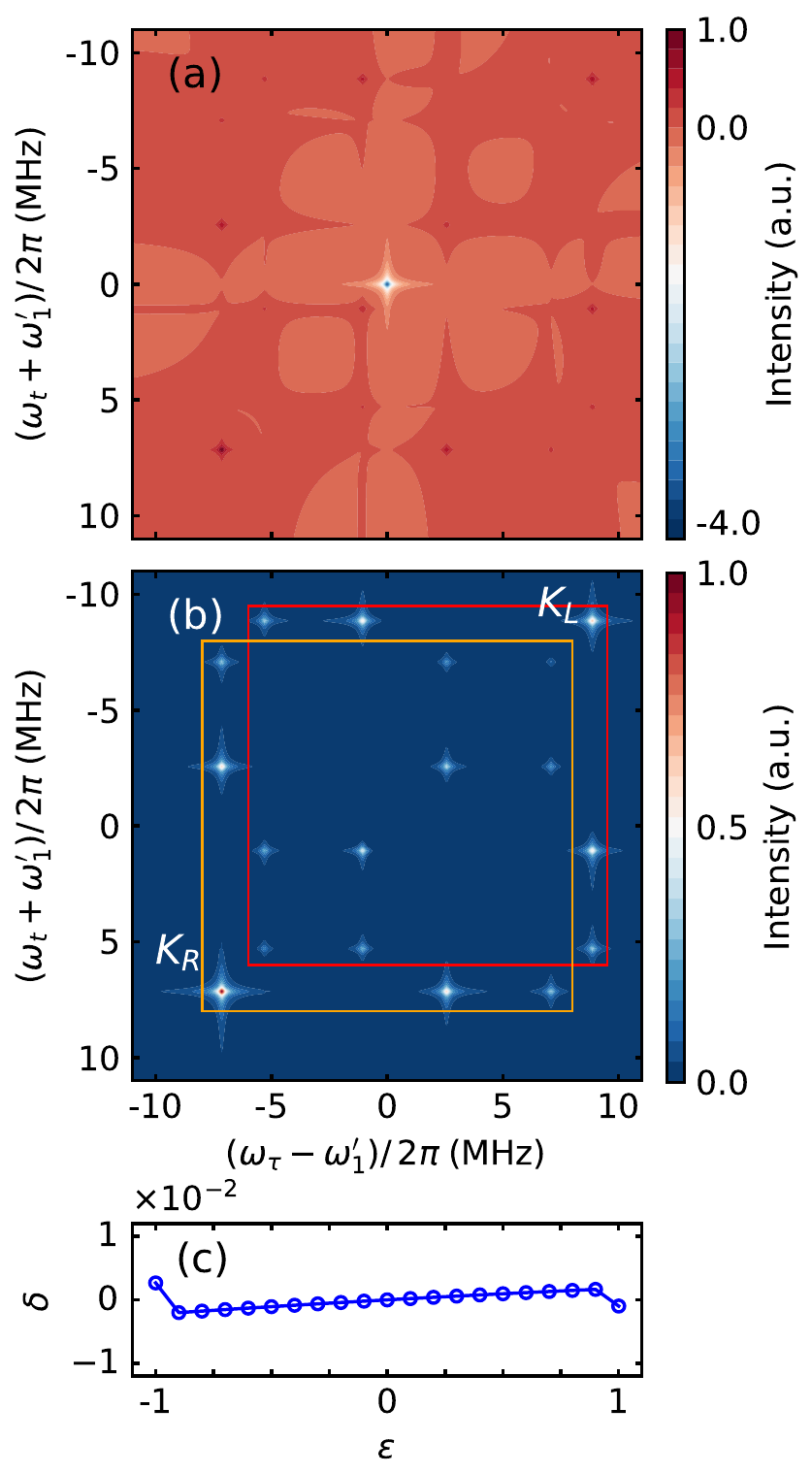}

\caption{(a) Chop detected 2D spectrum of the racemic mixture of gaseous 1,2-propanediol.
The deep negative peak at the center corresponds to the free eigenstate
$\ket{e_{1}^{\alpha}}$ of the 1,2-propanediol. (b) The truncated
2D spectrum of (a) by neglecting all the negative values. (c) The
error of the estimated enantiomeric excess with chop detection and
truncation. For the simulation, the parameters here are taken the
same with those in Fig.~\ref{2D-ThreePumping}. \label{2d-chopping}}

\end{figure}

Fig.~\ref{2d-chopping}(a) shows the chop detected 2D spectrum of
the racemic mixture. On the spectrum, a deep negative peak locates
at $(\omega_{1}',-\omega_{1}')$. This negative peak corresponds to
the subtraction of $\left|\tilde{\bm{A}}_{\mathrm{3e,\mathrm{off}}}^{\alpha}(\omega_{\tau},\omega_{t})\right|$
and strongly shades other chirality-dependent peaks. To eliminate
its disturbance, we make a truncation in $\tilde{\bm{A}}_{\mathrm{chop}}^{\alpha}(\omega_{\tau},\omega_{t})$
by neglecting all the negative values, and the result is presented
in Fig.~\ref{2d-chopping}(b). The truncated spectrum is almost the
same with the one in Fig.~\ref{2D-ThreePumping}(a) and still gives
good estimations of the enantiomeric excess with sufficiently low
errors, e.g., those below $1\times10^{-2}$ in Fig.~\ref{2d-chopping}(c).

\section{Summary and discussion}

This study has successfully demonstrated the effectiveness of the
enantiodetection method via 2D spectroscopy~\citep{Cai2022} under
more general experimental conditions. Previous method~\citep{Cai2022}
was proposed with a strict condition of one-photon resonance in the
CTLS and narrowband probe pulse assumption, limiting their applicability.
However, in this study, we have shown that even with three-photon
resonance in the CTLS and broadband probe pulses, different enantiomers
can still produce chirality-dependent peaks in the 2D spectrum. It
is important to note that identifying the chirality categorization
of the peaks is not possible in the 1D spectrum without further information.
Therefore, this method provides a significant advantage over traditional
techniques. Additionally, we have found that the peak intensities
for different enantiomers are not equal due to the inequality of the
transformation matrix under three-photon resonance. Therefore, the
enantiomeric excess of the chiral mixture is estimated using the racemic
mixture as the reference.

Real systems of chiral molecules contain massive redundant levels
that can disturb the chirality categorization. To eliminate this disturbance,
we have implemented chop detection and truncation. These methods have
proven to be effective in reducing the influence of redundant levels
and ensuring accurate enantiodetection results.

Overall, this study has provided valuable insights into the effectiveness
of the enantiodetection method via 2D spectroscopy and its applicability
under more general experimental conditions.
\begin{acknowledgments}
This work is supported by the National Natural Science Foundation
of China (Grants No. 12088101, No. 12074030, No. 12274107, No.12105011,
and No.U2230203) and the Research Funds of Hainan University {[}Grant
No. KYQD(ZR)23010{]}.
\end{acknowledgments}

\appendix

\section{Calculation of the final state and polarization without redundant
state}

In this appendix, we calculate the final state and polarization of
the system under three-photon resonance and with broadband probe pulses
inducing all three working transitions $\ket{g^{\alpha}}\leftrightarrow\ket{e_{j}^{\alpha}}$. 

For initial states $\ket{g^{\alpha}}$ or $\ket{e_{l}^{\alpha}}$
of the system, the evolved states after a free evolution time $s$
without the probe pulses are given in the Schr\"{o}dinger picture
as 
\begin{gather}
U_{\mathrm{cyc}}^{\alpha}(s)\ket{g^{\alpha}}=\ket{g^{\alpha}},\\
U_{\mathrm{cyc}}^{\alpha}(s)\ket{e_{l}^{\alpha}}=\sum_{j,j'}n_{j'j}^{\alpha}[n_{lj}^{\alpha}]^{*}e^{-iE_{j}^{\alpha}s}e^{-i\omega'_{j'}s}\ket{e_{j'}^{\alpha}}.
\end{gather}
On the other hand, the evolved states after a whole pulse duration
are given in the Schr\"{o}dinger picture as \begin{widetext}
\begin{gather}
U_{p,\mathrm{3e}}^{\alpha}\ket{g^{\alpha}}=\cos(\Omega_{p,\mathrm{3e}}^{\alpha}\delta t_{p})\ket{g^{\alpha}}-i\sin(\Omega_{p,\mathrm{3e}}^{\alpha}\delta t_{p})e^{i\vec{k}_{p}\cdot\vec{r}}\sum_{j}\frac{\Omega_{p,j}^{\alpha}}{\Omega_{p,\mathrm{3e}}^{\alpha}}\ket{e_{j}^{\alpha}},\label{evlvd-3e-1}\\
\begin{aligned}U_{p,\mathrm{3e}}^{\alpha}\ket{e_{j}^{\alpha}}= & -i\frac{\Omega_{p,j}^{\alpha}}{\Omega_{p,\mathrm{3e}}^{\alpha}}\sin(\Omega_{p,\mathrm{3e}}^{\alpha}\delta t_{p})e^{-i\vec{k}_{p}\cdot\vec{r}}\ket{g^{\alpha}}+\left[1-\left(\frac{\Omega_{p,j}^{\alpha}}{\Omega_{p,\mathrm{3e}}^{\alpha}}\right)^{2}\left(1-\cos(\Omega_{p,\mathrm{3e}}^{\alpha}\delta t_{p})\right)\right]\ket{e_{j}^{\alpha}}\\
+ & \sum_{l\neq j}\left[-\frac{\Omega_{p,j}^{\alpha}\Omega_{p,l}^{\alpha}}{(\Omega_{p,\mathrm{3e}}^{\alpha})^{2}}\left(1-\cos(\Omega_{p,\mathrm{3e}}^{\alpha}\delta t_{p})\right)\right]\ket{e_{l}^{\alpha}},
\end{aligned}
\label{evlvd-3e-2}
\end{gather}
\end{widetext}where $\Omega_{p,\mathrm{3e}}^{\alpha}=\sqrt{\Omega_{p,1}^{2}+\Omega_{p,2}^{2}+\Omega_{p,3}^{2}}$
and $U_{p,\mathrm{3e}}^{\alpha}=\exp[-iH_{0}^{\alpha}\delta t_{p}]\exp[-iV_{p,\mathrm{3e}}^{\alpha}\delta t_{p}]$
is the evolution operator within the pulse duration, while, however,
the constant phase factors introduced by $\exp[-iH_{0}^{\alpha}\delta t_{p}]$
is neglected in Eqs.~(\ref{evlvd-3e-1}) and (\ref{evlvd-3e-2}).
In the perturbative regime, i.e., $\left|\Omega_{p,\mathrm{3e}}^{\alpha}\delta t_{p}\right|\ll1$,
the evolved states are approximately
\begin{gather}
U_{p,\mathrm{3e}}^{\alpha}\ket{g^{\alpha}}=\mathcal{N}_{p,0}^{\alpha}\left(\ket{g^{\alpha}}+e^{i\vec{k}_{p}\cdot\vec{r}}\sum_{j}\beta_{p,j}^{\alpha}\ket{e_{j}^{\alpha}}\right),\\
U_{p,\mathrm{3e}}^{\alpha}\ket{e_{j}^{\alpha}}=\mathcal{N}_{p,j}^{\alpha}\left(\ket{e_{j}^{\alpha}}+e^{-i\vec{k}_{p}\cdot\vec{r}}\beta_{p,j}^{\alpha}\ket{g^{\alpha}}\right).
\end{gather}

In our proposed experiment with the sequence of two probe pulses,
the final state of the system with initial state $\ket{\psi_{0}^{\alpha}}=\ket{g^{\alpha}}$
is calculated as\begin{widetext}
\begin{equation}
\begin{aligned}\ket{\psi_{\mathrm{3e}}^{\alpha}(\tau,t)} & =U_{\mathrm{cyc}}^{\alpha}(t)U_{b,\mathrm{3e}}^{\alpha}U_{\mathrm{cyc}}^{\alpha}(\tau)U_{a,\mathrm{3e}}^{\alpha}\ket{\psi_{0}^{\alpha}}\\
 & =\mathcal{N}_{a,0}^{\alpha}\mathcal{N}_{b,0}^{\alpha}e^{i\vec{k}_{b}\cdot\vec{r}}\sum_{m}\beta_{b,m}^{\alpha}\sum_{j,l}n_{lj}^{\alpha}[n_{mj}^{\alpha}]^{*}e^{-iE_{j}^{\alpha}t}e^{-i\omega_{l}'t}\ket{e_{l}^{\alpha}}\\
 & +\mathcal{N}_{a,0}^{\alpha}e^{i(\vec{k}_{a}-\vec{k}_{b})\cdot\vec{r}}\sum_{m}\beta_{a,m}^{\alpha}\sum_{j,l}n_{lj}^{\alpha}[n_{mj}^{\alpha}]^{*}e^{-iE_{j}^{\alpha}\tau}e^{-i\omega_{l}'\tau}\mathcal{N}_{b,l}^{\alpha}\beta_{b,l}^{\alpha}\ket{g^{\alpha}}\\
 & +\mathcal{N}_{a,0}^{\alpha}\mathcal{N}_{b,0}^{\alpha}\ket{g^{\alpha}}+\mathcal{N}_{a,0}^{\alpha}e^{i\vec{k}_{a}\cdot\vec{r}}\sum_{m}\beta_{a,m}^{\alpha}\sum_{j,l}n_{lj}^{\alpha}[n_{mj}^{\alpha}]^{*}e^{-iE_{j}^{\alpha}\tau}e^{-i\omega_{l}'\tau}\mathcal{N}_{b,l}^{\alpha}\sum_{j',l'}n_{l'j'}^{\alpha}[n_{lj'}^{\alpha}]^{*}e^{-iE_{j'}^{\alpha}t}e^{-i\omega_{l'}'t}\ket{e_{l'}^{\alpha}}.
\end{aligned}
\end{equation}
\end{widetext}With the final state, the polarization after the delay
time $\tau$ and data collecting time $t$ is given by $\bm{P}_{\mathrm{3e}}^{\alpha}(\tau,t)=\bra{\psi_{\mathrm{3e}}^{\alpha}(\tau,t)}\hat{\bm{\mu}}^{\alpha}\ket{\psi_{\mathrm{3e}}^{\alpha}(\tau,t)}$.
Adding the decoherence during $\tau$ and $t$, the amplitude of the
rephasing part of the final polarization with phase factor $e^{i(-\vec{k}_{a}+2\vec{k}_{b})\cdot\vec{r}}$
is\begin{widetext}
\begin{equation}
\bm{A}_{\mathrm{3e}}^{\alpha}(\tau,t)=\left[\mathcal{N}_{a,0}^{\alpha}\right]^{2}\mathcal{N}_{b,0}^{\alpha}\sum_{m,m'}\left[\beta_{a,m}^{\alpha}\right]^{*}\beta_{b,m'}^{\alpha}\sum_{j,l}[n_{lj}^{\alpha}]^{*}n_{mj}^{\alpha}e^{i(\omega_{l}'+E_{j}^{\alpha})\tau}e^{-\Gamma\tau}\mathcal{N}_{b,l}^{\alpha}\left[\beta_{b,l}^{\alpha}\right]^{*}\sum_{j',l'}n_{l'j'}^{\alpha}[n_{m'j'}^{\alpha}]^{*}e^{-i(\omega_{l'}'+E_{j'}^{\alpha})t}e^{-\Gamma t}\bm{\mu}_{0l'}^{\alpha},\label{sig-rp-3pump-appdx}
\end{equation}
\end{widetext}which is identical with the one in Eq.~(\ref{sig-rp-3pump}).
Letting $\bm{\mu}_{02}^{\alpha}$ and $\bm{\mu}_{03}^{\alpha}$ to
be zero, i.e., only the transition $\ket{g^{\alpha}}\leftrightarrow\ket{e_{1}^{\alpha}}$
can be induced by the probe pulses, Eq.~(\ref{sig-rp-3pump-appdx})
thus matches Eq.~(\ref{sig-rp1}).

\bibliographystyle{apsrev4-2}
\phantomsection\addcontentsline{toc}{section}{\refname}\bibliography{ExtendedChiral2D_bib}

\begin{thebibliography}{44}%
\makeatletter
\providecommand \@ifxundefined [1]{%
 \@ifx{#1\undefined}
}%
\providecommand \@ifnum [1]{%
 \ifnum #1\expandafter \@firstoftwo
 \else \expandafter \@secondoftwo
 \fi
}%
\providecommand \@ifx [1]{%
 \ifx #1\expandafter \@firstoftwo
 \else \expandafter \@secondoftwo
 \fi
}%
\providecommand \natexlab [1]{#1}%
\providecommand \enquote  [1]{``#1''}%
\providecommand \bibnamefont  [1]{#1}%
\providecommand \bibfnamefont [1]{#1}%
\providecommand \citenamefont [1]{#1}%
\providecommand \href@noop [0]{\@secondoftwo}%
\providecommand \href [0]{\begingroup \@sanitize@url \@href}%
\providecommand \@href[1]{\@@startlink{#1}\@@href}%
\providecommand \@@href[1]{\endgroup#1\@@endlink}%
\providecommand \@sanitize@url [0]{\catcode `\\12\catcode `\$12\catcode
  `\&12\catcode `\#12\catcode `\^12\catcode `\_12\catcode `\%12\relax}%
\providecommand \@@startlink[1]{}%
\providecommand \@@endlink[0]{}%
\providecommand \url  [0]{\begingroup\@sanitize@url \@url }%
\providecommand \@url [1]{\endgroup\@href {#1}{\urlprefix }}%
\providecommand \urlprefix  [0]{URL }%
\providecommand \Eprint [0]{\href }%
\providecommand \doibase [0]{https://doi.org/}%
\providecommand \selectlanguage [0]{\@gobble}%
\providecommand \bibinfo  [0]{\@secondoftwo}%
\providecommand \bibfield  [0]{\@secondoftwo}%
\providecommand \translation [1]{[#1]}%
\providecommand \BibitemOpen [0]{}%
\providecommand \bibitemStop [0]{}%
\providecommand \bibitemNoStop [0]{.\EOS\space}%
\providecommand \EOS [0]{\spacefactor3000\relax}%
\providecommand \BibitemShut  [1]{\csname bibitem#1\endcsname}%
\let\auto@bib@innerbib\@empty
\bibitem [{\citenamefont {Quack}(1989)}]{Quack1989}%
  \BibitemOpen
  \bibfield  {author} {\bibinfo {author} {\bibfnamefont {M.}~\bibnamefont
  {Quack}},\ }\href {https://doi.org/https://doi.org/10.1002/anie.198905711}
  {\bibfield  {journal} {\bibinfo  {journal} {Angew. Chem. Int. Ed.}\ }\textbf
  {\bibinfo {volume} {28}},\ \bibinfo {pages} {571} (\bibinfo {year}
  {1989})}\BibitemShut {NoStop}%
\bibitem [{\citenamefont {Li}\ \emph {et~al.}(2007)\citenamefont {Li},
  \citenamefont {Bruder},\ and\ \citenamefont {Sun}}]{LiYong2007}%
  \BibitemOpen
  \bibfield  {author} {\bibinfo {author} {\bibfnamefont {Y.}~\bibnamefont
  {Li}}, \bibinfo {author} {\bibfnamefont {C.}~\bibnamefont {Bruder}},\ and\
  \bibinfo {author} {\bibfnamefont {C.~P.}\ \bibnamefont {Sun}},\ }\href
  {https://doi.org/10.1103/PhysRevLett.99.130403} {\bibfield  {journal}
  {\bibinfo  {journal} {Phys. Rev. Lett.}\ }\textbf {\bibinfo {volume} {99}},\
  \bibinfo {pages} {130403} (\bibinfo {year} {2007})}\BibitemShut {NoStop}%
\bibitem [{\citenamefont {Li}\ and\ \citenamefont
  {Shapiro}(2010)}]{LiXuan2010}%
  \BibitemOpen
  \bibfield  {author} {\bibinfo {author} {\bibfnamefont {X.}~\bibnamefont
  {Li}}\ and\ \bibinfo {author} {\bibfnamefont {M.}~\bibnamefont {Shapiro}},\
  }\href {https://doi.org/10.1063/1.3429884} {\bibfield  {journal} {\bibinfo
  {journal} {J. Chem. Phys.}\ }\textbf {\bibinfo {volume} {132}},\ \bibinfo
  {pages} {194315} (\bibinfo {year} {2010})}\BibitemShut {NoStop}%
\bibitem [{\citenamefont {Suzuki}\ \emph {et~al.}(2019)\citenamefont {Suzuki},
  \citenamefont {Momose},\ and\ \citenamefont {Buhmann}}]{Buhmann2019}%
  \BibitemOpen
  \bibfield  {author} {\bibinfo {author} {\bibfnamefont {F.}~\bibnamefont
  {Suzuki}}, \bibinfo {author} {\bibfnamefont {T.}~\bibnamefont {Momose}},\
  and\ \bibinfo {author} {\bibfnamefont {S.~Y.}\ \bibnamefont {Buhmann}},\
  }\href {https://doi.org/10.1103/PhysRevA.99.012513} {\bibfield  {journal}
  {\bibinfo  {journal} {Phys. Rev. A}\ }\textbf {\bibinfo {volume} {99}},\
  \bibinfo {pages} {012513} (\bibinfo {year} {2019})}\BibitemShut {NoStop}%
\bibitem [{\citenamefont {Liu}\ \emph {et~al.}(2021)\citenamefont {Liu},
  \citenamefont {Ye}, \citenamefont {Sun},\ and\ \citenamefont
  {Li}}]{LiuBo2021}%
  \BibitemOpen
  \bibfield  {author} {\bibinfo {author} {\bibfnamefont {B.}~\bibnamefont
  {Liu}}, \bibinfo {author} {\bibfnamefont {C.}~\bibnamefont {Ye}}, \bibinfo
  {author} {\bibfnamefont {C.~P.}\ \bibnamefont {Sun}},\ and\ \bibinfo {author}
  {\bibfnamefont {Y.}~\bibnamefont {Li}},\ }\href
  {https://doi.org/10.1103/PhysRevA.104.013113} {\bibfield  {journal} {\bibinfo
   {journal} {Phys. Rev. A}\ }\textbf {\bibinfo {volume} {104}},\ \bibinfo
  {pages} {013113} (\bibinfo {year} {2021})}\BibitemShut {NoStop}%
\bibitem [{\citenamefont {Shapiro}\ \emph {et~al.}(2000)\citenamefont
  {Shapiro}, \citenamefont {Frishman},\ and\ \citenamefont
  {Brumer}}]{Shapiro2000}%
  \BibitemOpen
  \bibfield  {author} {\bibinfo {author} {\bibfnamefont {M.}~\bibnamefont
  {Shapiro}}, \bibinfo {author} {\bibfnamefont {E.}~\bibnamefont {Frishman}},\
  and\ \bibinfo {author} {\bibfnamefont {P.}~\bibnamefont {Brumer}},\ }\href
  {https://doi.org/10.1103/PhysRevLett.84.1669} {\bibfield  {journal} {\bibinfo
   {journal} {Phys. Rev. Lett.}\ }\textbf {\bibinfo {volume} {84}},\ \bibinfo
  {pages} {1669} (\bibinfo {year} {2000})}\BibitemShut {NoStop}%
\bibitem [{\citenamefont {Kr\'al}\ \emph {et~al.}(2003)\citenamefont {Kr\'al},
  \citenamefont {Thanopulos}, \citenamefont {Shapiro},\ and\ \citenamefont
  {Cohen}}]{Kral2003}%
  \BibitemOpen
  \bibfield  {author} {\bibinfo {author} {\bibfnamefont {P.}~\bibnamefont
  {Kr\'al}}, \bibinfo {author} {\bibfnamefont {I.}~\bibnamefont {Thanopulos}},
  \bibinfo {author} {\bibfnamefont {M.}~\bibnamefont {Shapiro}},\ and\ \bibinfo
  {author} {\bibfnamefont {D.}~\bibnamefont {Cohen}},\ }\href
  {https://doi.org/10.1103/PhysRevLett.90.033001} {\bibfield  {journal}
  {\bibinfo  {journal} {Phys. Rev. Lett.}\ }\textbf {\bibinfo {volume} {90}},\
  \bibinfo {pages} {033001} (\bibinfo {year} {2003})}\BibitemShut {NoStop}%
\bibitem [{\citenamefont {Ye}\ \emph {et~al.}(2021{\natexlab{a}})\citenamefont
  {Ye}, \citenamefont {Chen}, \citenamefont {Zhang},\ and\ \citenamefont
  {Li}}]{YeChong2021JPB}%
  \BibitemOpen
  \bibfield  {author} {\bibinfo {author} {\bibfnamefont {C.}~\bibnamefont
  {Ye}}, \bibinfo {author} {\bibfnamefont {Y.-Y.}\ \bibnamefont {Chen}},
  \bibinfo {author} {\bibfnamefont {Q.}~\bibnamefont {Zhang}},\ and\ \bibinfo
  {author} {\bibfnamefont {Y.}~\bibnamefont {Li}},\ }\href
  {https://doi.org/10.1088/1361-6455/ac09c3} {\bibfield  {journal} {\bibinfo
  {journal} {J. Phys. B: At. Mol. Opt. Phys.}\ }\textbf {\bibinfo {volume}
  {54}},\ \bibinfo {pages} {145102} (\bibinfo {year}
  {2021}{\natexlab{a}})}\BibitemShut {NoStop}%
\bibitem [{\citenamefont {Ye}\ \emph {et~al.}(2021{\natexlab{b}})\citenamefont
  {Ye}, \citenamefont {Liu}, \citenamefont {Chen},\ and\ \citenamefont
  {Li}}]{YeChong2021PRA}%
  \BibitemOpen
  \bibfield  {author} {\bibinfo {author} {\bibfnamefont {C.}~\bibnamefont
  {Ye}}, \bibinfo {author} {\bibfnamefont {B.}~\bibnamefont {Liu}}, \bibinfo
  {author} {\bibfnamefont {Y.-Y.}\ \bibnamefont {Chen}},\ and\ \bibinfo
  {author} {\bibfnamefont {Y.}~\bibnamefont {Li}},\ }\href
  {https://doi.org/10.1103/PhysRevA.103.022830} {\bibfield  {journal} {\bibinfo
   {journal} {Phys. Rev. A}\ }\textbf {\bibinfo {volume} {103}},\ \bibinfo
  {pages} {022830} (\bibinfo {year} {2021}{\natexlab{b}})}\BibitemShut
  {NoStop}%
\bibitem [{\citenamefont {He}\ \emph {et~al.}(2011)\citenamefont {He},
  \citenamefont {Wang}, \citenamefont {Dukor},\ and\ \citenamefont
  {Nafie}}]{HeYanan2011}%
  \BibitemOpen
  \bibfield  {author} {\bibinfo {author} {\bibfnamefont {Y.}~\bibnamefont
  {He}}, \bibinfo {author} {\bibfnamefont {B.}~\bibnamefont {Wang}}, \bibinfo
  {author} {\bibfnamefont {R.~K.}\ \bibnamefont {Dukor}},\ and\ \bibinfo
  {author} {\bibfnamefont {L.~A.}\ \bibnamefont {Nafie}},\ }\href
  {https://doi.org/10.1366/11-06321} {\bibfield  {journal} {\bibinfo  {journal}
  {Appl. Spectrosc.}\ }\textbf {\bibinfo {volume} {65}},\ \bibinfo {pages}
  {699} (\bibinfo {year} {2011})}\BibitemShut {NoStop}%
\bibitem [{\citenamefont {Stephens}(1985)}]{Stephens1985}%
  \BibitemOpen
  \bibfield  {author} {\bibinfo {author} {\bibfnamefont {P.~J.}\ \bibnamefont
  {Stephens}},\ }\href {https://doi.org/10.1021/j100251a006} {\bibfield
  {journal} {\bibinfo  {journal} {J. Phys. Chem.}\ }\textbf {\bibinfo {volume}
  {89}},\ \bibinfo {pages} {748} (\bibinfo {year} {1985})}\BibitemShut
  {NoStop}%
\bibitem [{\citenamefont {Begzjav}\ \emph {et~al.}(2019)\citenamefont
  {Begzjav}, \citenamefont {Zhang}, \citenamefont {Scully},\ and\ \citenamefont
  {Agarwal}}]{Begzjav2019}%
  \BibitemOpen
  \bibfield  {author} {\bibinfo {author} {\bibfnamefont {T.~K.}\ \bibnamefont
  {Begzjav}}, \bibinfo {author} {\bibfnamefont {Z.}~\bibnamefont {Zhang}},
  \bibinfo {author} {\bibfnamefont {M.~O.}\ \bibnamefont {Scully}},\ and\
  \bibinfo {author} {\bibfnamefont {G.~S.}\ \bibnamefont {Agarwal}},\ }\href
  {https://doi.org/10.1364/OE.27.013965} {\bibfield  {journal} {\bibinfo
  {journal} {Opt. Express}\ }\textbf {\bibinfo {volume} {27}},\ \bibinfo
  {pages} {13965} (\bibinfo {year} {2019})}\BibitemShut {NoStop}%
\bibitem [{\citenamefont {Fanood}\ \emph {et~al.}(2015)\citenamefont {Fanood},
  \citenamefont {Ram}, \citenamefont {Lehmann}, \citenamefont {Powis},\ and\
  \citenamefont {Janssen}}]{Fanood2015}%
  \BibitemOpen
  \bibfield  {author} {\bibinfo {author} {\bibfnamefont {M.~M.~R.}\
  \bibnamefont {Fanood}}, \bibinfo {author} {\bibfnamefont {N.~B.}\
  \bibnamefont {Ram}}, \bibinfo {author} {\bibfnamefont {C.~S.}\ \bibnamefont
  {Lehmann}}, \bibinfo {author} {\bibfnamefont {I.}~\bibnamefont {Powis}},\
  and\ \bibinfo {author} {\bibfnamefont {M.~H.~M.}\ \bibnamefont {Janssen}},\
  }\href {https://doi.org/10.1038/ncomms8511} {\bibfield  {journal} {\bibinfo
  {journal} {Nature Communications}\ }\textbf {\bibinfo {volume} {6}},\
  \bibinfo {pages} {7511} (\bibinfo {year} {2015})}\BibitemShut {NoStop}%
\bibitem [{\citenamefont {Ghosh}\ and\ \citenamefont
  {Fischer}(2006)}]{Ghosh2006}%
  \BibitemOpen
  \bibfield  {author} {\bibinfo {author} {\bibfnamefont {A.}~\bibnamefont
  {Ghosh}}\ and\ \bibinfo {author} {\bibfnamefont {P.}~\bibnamefont
  {Fischer}},\ }\href {https://doi.org/10.1103/PhysRevLett.97.173002}
  {\bibfield  {journal} {\bibinfo  {journal} {Phys. Rev. Lett.}\ }\textbf
  {\bibinfo {volume} {97}},\ \bibinfo {pages} {173002} (\bibinfo {year}
  {2006})}\BibitemShut {NoStop}%
\bibitem [{\citenamefont {Patterson}\ \emph {et~al.}(2013)\citenamefont
  {Patterson}, \citenamefont {Schnell},\ and\ \citenamefont
  {Doyle}}]{Patterson2013Nature}%
  \BibitemOpen
  \bibfield  {author} {\bibinfo {author} {\bibfnamefont {D.}~\bibnamefont
  {Patterson}}, \bibinfo {author} {\bibfnamefont {M.}~\bibnamefont {Schnell}},\
  and\ \bibinfo {author} {\bibfnamefont {J.~M.}\ \bibnamefont {Doyle}},\ }\href
  {https://doi.org/10.1038/nature12150} {\bibfield  {journal} {\bibinfo
  {journal} {Nature}\ }\textbf {\bibinfo {volume} {497}},\ \bibinfo {pages}
  {475} (\bibinfo {year} {2013})}\BibitemShut {NoStop}%
\bibitem [{\citenamefont {Patterson}\ and\ \citenamefont
  {Doyle}(2013)}]{Patterson2013PRL}%
  \BibitemOpen
  \bibfield  {author} {\bibinfo {author} {\bibfnamefont {D.}~\bibnamefont
  {Patterson}}\ and\ \bibinfo {author} {\bibfnamefont {J.~M.}\ \bibnamefont
  {Doyle}},\ }\href {https://doi.org/10.1103/PhysRevLett.111.023008} {\bibfield
   {journal} {\bibinfo  {journal} {Phys. Rev. Lett.}\ }\textbf {\bibinfo
  {volume} {111}},\ \bibinfo {pages} {023008} (\bibinfo {year}
  {2013})}\BibitemShut {NoStop}%
\bibitem [{\citenamefont {Lobsiger}\ \emph {et~al.}(2015)\citenamefont
  {Lobsiger}, \citenamefont {Perez}, \citenamefont {Evangelisti}, \citenamefont
  {Lehmann},\ and\ \citenamefont {Pate}}]{Lobsiger2015}%
  \BibitemOpen
  \bibfield  {author} {\bibinfo {author} {\bibfnamefont {S.}~\bibnamefont
  {Lobsiger}}, \bibinfo {author} {\bibfnamefont {C.}~\bibnamefont {Perez}},
  \bibinfo {author} {\bibfnamefont {L.}~\bibnamefont {Evangelisti}}, \bibinfo
  {author} {\bibfnamefont {K.~K.}\ \bibnamefont {Lehmann}},\ and\ \bibinfo
  {author} {\bibfnamefont {B.~H.}\ \bibnamefont {Pate}},\ }\href
  {https://doi.org/10.1021/jz502312t} {\bibfield  {journal} {\bibinfo
  {journal} {J. Phys. Chem. Lett.}\ }\textbf {\bibinfo {volume} {6}},\ \bibinfo
  {pages} {196} (\bibinfo {year} {2015})}\BibitemShut {NoStop}%
\bibitem [{\citenamefont {Shubert}\ \emph {et~al.}(2016)\citenamefont
  {Shubert}, \citenamefont {Schmitz}, \citenamefont {Pérez}, \citenamefont
  {Medcraft}, \citenamefont {Krin}, \citenamefont {Domingos}, \citenamefont
  {Patterson},\ and\ \citenamefont {Schnell}}]{Shubert2016}%
  \BibitemOpen
  \bibfield  {author} {\bibinfo {author} {\bibfnamefont {V.~A.}\ \bibnamefont
  {Shubert}}, \bibinfo {author} {\bibfnamefont {D.}~\bibnamefont {Schmitz}},
  \bibinfo {author} {\bibfnamefont {C.}~\bibnamefont {Pérez}}, \bibinfo
  {author} {\bibfnamefont {C.}~\bibnamefont {Medcraft}}, \bibinfo {author}
  {\bibfnamefont {A.}~\bibnamefont {Krin}}, \bibinfo {author} {\bibfnamefont
  {S.~R.}\ \bibnamefont {Domingos}}, \bibinfo {author} {\bibfnamefont
  {D.}~\bibnamefont {Patterson}},\ and\ \bibinfo {author} {\bibfnamefont
  {M.}~\bibnamefont {Schnell}},\ }\href
  {https://doi.org/10.1021/acs.jpclett.5b02443} {\bibfield  {journal} {\bibinfo
   {journal} {J. Phys. Chem. Lett.}\ }\textbf {\bibinfo {volume} {7}},\
  \bibinfo {pages} {341} (\bibinfo {year} {2016})}\BibitemShut {NoStop}%
\bibitem [{\citenamefont {Ye}\ \emph {et~al.}(2019)\citenamefont {Ye},
  \citenamefont {Zhang}, \citenamefont {Chen},\ and\ \citenamefont
  {Li}}]{YeChong2019}%
  \BibitemOpen
  \bibfield  {author} {\bibinfo {author} {\bibfnamefont {C.}~\bibnamefont
  {Ye}}, \bibinfo {author} {\bibfnamefont {Q.}~\bibnamefont {Zhang}}, \bibinfo
  {author} {\bibfnamefont {Y.-Y.}\ \bibnamefont {Chen}},\ and\ \bibinfo
  {author} {\bibfnamefont {Y.}~\bibnamefont {Li}},\ }\href
  {https://doi.org/10.1103/PhysRevA.100.033411} {\bibfield  {journal} {\bibinfo
   {journal} {Phys. Rev. A}\ }\textbf {\bibinfo {volume} {100}},\ \bibinfo
  {pages} {033411} (\bibinfo {year} {2019})}\BibitemShut {NoStop}%
\bibitem [{\citenamefont {Cai}\ \emph {et~al.}(2022)\citenamefont {Cai},
  \citenamefont {Ye}, \citenamefont {Dong},\ and\ \citenamefont
  {Li}}]{Cai2022}%
  \BibitemOpen
  \bibfield  {author} {\bibinfo {author} {\bibfnamefont {M.-R.}\ \bibnamefont
  {Cai}}, \bibinfo {author} {\bibfnamefont {C.}~\bibnamefont {Ye}}, \bibinfo
  {author} {\bibfnamefont {H.}~\bibnamefont {Dong}},\ and\ \bibinfo {author}
  {\bibfnamefont {Y.}~\bibnamefont {Li}},\ }\href
  {https://doi.org/10.1103/PhysRevLett.129.103201} {\bibfield  {journal}
  {\bibinfo  {journal} {Phys. Rev. Lett.}\ }\textbf {\bibinfo {volume} {129}},\
  \bibinfo {pages} {103201} (\bibinfo {year} {2022})}\BibitemShut {NoStop}%
\bibitem [{\citenamefont {Jia}\ and\ \citenamefont {Wei}(2011)}]{Jia2011}%
  \BibitemOpen
  \bibfield  {author} {\bibinfo {author} {\bibfnamefont {W.~Z.}\ \bibnamefont
  {Jia}}\ and\ \bibinfo {author} {\bibfnamefont {L.~F.}\ \bibnamefont {Wei}},\
  }\href {https://doi.org/10.1103/PhysRevA.84.053849} {\bibfield  {journal}
  {\bibinfo  {journal} {Phys. Rev. A}\ }\textbf {\bibinfo {volume} {84}},\
  \bibinfo {pages} {053849} (\bibinfo {year} {2011})}\BibitemShut {NoStop}%
\bibitem [{\citenamefont {Ye}\ \emph {et~al.}(2021{\natexlab{c}})\citenamefont
  {Ye}, \citenamefont {Sun},\ and\ \citenamefont {Zhang}}]{YeChong2021JPCL}%
  \BibitemOpen
  \bibfield  {author} {\bibinfo {author} {\bibfnamefont {C.}~\bibnamefont
  {Ye}}, \bibinfo {author} {\bibfnamefont {Y.}~\bibnamefont {Sun}},\ and\
  \bibinfo {author} {\bibfnamefont {X.}~\bibnamefont {Zhang}},\ }\href
  {https://doi.org/10.1021/acs.jpclett.1c02196} {\bibfield  {journal} {\bibinfo
   {journal} {J. Phys. Chem. Lett.}\ }\textbf {\bibinfo {volume} {12}},\
  \bibinfo {pages} {8591} (\bibinfo {year} {2021}{\natexlab{c}})}\BibitemShut
  {NoStop}%
\bibitem [{\citenamefont {Kr\'al}\ and\ \citenamefont
  {Shapiro}(2001)}]{Kral2001}%
  \BibitemOpen
  \bibfield  {author} {\bibinfo {author} {\bibfnamefont {P.}~\bibnamefont
  {Kr\'al}}\ and\ \bibinfo {author} {\bibfnamefont {M.}~\bibnamefont
  {Shapiro}},\ }\href {https://doi.org/10.1103/PhysRevLett.87.183002}
  {\bibfield  {journal} {\bibinfo  {journal} {Phys. Rev. Lett.}\ }\textbf
  {\bibinfo {volume} {87}},\ \bibinfo {pages} {183002} (\bibinfo {year}
  {2001})}\BibitemShut {NoStop}%
\bibitem [{\citenamefont {Jacob}\ and\ \citenamefont
  {Hornberger}(2012)}]{Jacob2012}%
  \BibitemOpen
  \bibfield  {author} {\bibinfo {author} {\bibfnamefont {A.}~\bibnamefont
  {Jacob}}\ and\ \bibinfo {author} {\bibfnamefont {K.}~\bibnamefont
  {Hornberger}},\ }\href {https://doi.org/10.1063/1.4738753} {\bibfield
  {journal} {\bibinfo  {journal} {J. Chem. Phys.}\ }\textbf {\bibinfo {volume}
  {137}},\ \bibinfo {pages} {044313} (\bibinfo {year} {2012})}\BibitemShut
  {NoStop}%
\bibitem [{\citenamefont {Lehmann}(2018)}]{Lehmann2018}%
  \BibitemOpen
  \bibfield  {author} {\bibinfo {author} {\bibfnamefont {K.~K.}\ \bibnamefont
  {Lehmann}},\ }\href {https://doi.org/10.1063/1.5045052} {\bibfield  {journal}
  {\bibinfo  {journal} {J. Chem. Phys.}\ }\textbf {\bibinfo {volume} {149}},\
  \bibinfo {pages} {094201} (\bibinfo {year} {2018})}\BibitemShut {NoStop}%
\bibitem [{\citenamefont {Ye}\ \emph {et~al.}(2018)\citenamefont {Ye},
  \citenamefont {Zhang},\ and\ \citenamefont {Li}}]{YeChong2018}%
  \BibitemOpen
  \bibfield  {author} {\bibinfo {author} {\bibfnamefont {C.}~\bibnamefont
  {Ye}}, \bibinfo {author} {\bibfnamefont {Q.}~\bibnamefont {Zhang}},\ and\
  \bibinfo {author} {\bibfnamefont {Y.}~\bibnamefont {Li}},\ }\href
  {https://doi.org/10.1103/PhysRevA.98.063401} {\bibfield  {journal} {\bibinfo
  {journal} {Phys. Rev. A}\ }\textbf {\bibinfo {volume} {98}},\ \bibinfo
  {pages} {063401} (\bibinfo {year} {2018})}\BibitemShut {NoStop}%
\bibitem [{\citenamefont {Leibscher}\ \emph {et~al.}(2019)\citenamefont
  {Leibscher}, \citenamefont {Giesen},\ and\ \citenamefont
  {Koch}}]{Leibscher2019}%
  \BibitemOpen
  \bibfield  {author} {\bibinfo {author} {\bibfnamefont {M.}~\bibnamefont
  {Leibscher}}, \bibinfo {author} {\bibfnamefont {T.~F.}\ \bibnamefont
  {Giesen}},\ and\ \bibinfo {author} {\bibfnamefont {C.~P.}\ \bibnamefont
  {Koch}},\ }\href {https://doi.org/10.1063/1.5097406} {\bibfield  {journal}
  {\bibinfo  {journal} {J. Chem. Phys.}\ }\textbf {\bibinfo {volume} {151}},\
  \bibinfo {pages} {014302} (\bibinfo {year} {2019})}\BibitemShut {NoStop}%
\bibitem [{\citenamefont {Zhang}\ \emph {et~al.}(2020)\citenamefont {Zhang},
  \citenamefont {Chen}, \citenamefont {Ye},\ and\ \citenamefont
  {Li}}]{ZhangQS2020}%
  \BibitemOpen
  \bibfield  {author} {\bibinfo {author} {\bibfnamefont {Q.}~\bibnamefont
  {Zhang}}, \bibinfo {author} {\bibfnamefont {Y.-Y.}\ \bibnamefont {Chen}},
  \bibinfo {author} {\bibfnamefont {C.}~\bibnamefont {Ye}},\ and\ \bibinfo
  {author} {\bibfnamefont {Y.}~\bibnamefont {Li}},\ }\href
  {https://doi.org/10.1088/1361-6455/abc143} {\bibfield  {journal} {\bibinfo
  {journal} {J. Phys. B: At. Mol. Opt. Phys.}\ }\textbf {\bibinfo {volume}
  {53}},\ \bibinfo {pages} {235103} (\bibinfo {year} {2020})}\BibitemShut
  {NoStop}%
\bibitem [{\citenamefont {Jonas}(2003)}]{Jonas2003}%
  \BibitemOpen
  \bibfield  {author} {\bibinfo {author} {\bibfnamefont {D.~M.}\ \bibnamefont
  {Jonas}},\ }\href {https://doi.org/10.1126/science.1085687} {\bibfield
  {journal} {\bibinfo  {journal} {Science}\ }\textbf {\bibinfo {volume}
  {300}},\ \bibinfo {pages} {1515} (\bibinfo {year} {2003})}\BibitemShut
  {NoStop}%
\bibitem [{\citenamefont {Vitanov}\ and\ \citenamefont
  {Drewsen}(2019)}]{Vitanov2019}%
  \BibitemOpen
  \bibfield  {author} {\bibinfo {author} {\bibfnamefont {N.~V.}\ \bibnamefont
  {Vitanov}}\ and\ \bibinfo {author} {\bibfnamefont {M.}~\bibnamefont
  {Drewsen}},\ }\href {https://doi.org/10.1103/PhysRevLett.122.173202}
  {\bibfield  {journal} {\bibinfo  {journal} {Phys. Rev. Lett.}\ }\textbf
  {\bibinfo {volume} {122}},\ \bibinfo {pages} {173202} (\bibinfo {year}
  {2019})}\BibitemShut {NoStop}%
\bibitem [{\citenamefont {Reimann}\ \emph {et~al.}(2021)\citenamefont
  {Reimann}, \citenamefont {Woerner},\ and\ \citenamefont
  {Elsaesser}}]{Reimann2021}%
  \BibitemOpen
  \bibfield  {author} {\bibinfo {author} {\bibfnamefont {K.}~\bibnamefont
  {Reimann}}, \bibinfo {author} {\bibfnamefont {M.}~\bibnamefont {Woerner}},\
  and\ \bibinfo {author} {\bibfnamefont {T.}~\bibnamefont {Elsaesser}},\ }\href
  {https://doi.org/10.1063/5.0046664} {\bibfield  {journal} {\bibinfo
  {journal} {J. Chem. Phys.}\ }\textbf {\bibinfo {volume} {154}},\ \bibinfo
  {pages} {120901} (\bibinfo {year} {2021})}\BibitemShut {NoStop}%
\bibitem [{\citenamefont {Woerner}\ \emph {et~al.}(2013)\citenamefont
  {Woerner}, \citenamefont {Kuehn}, \citenamefont {Bowlan}, \citenamefont
  {Reimann},\ and\ \citenamefont {Elsaesser}}]{Woerner2013}%
  \BibitemOpen
  \bibfield  {author} {\bibinfo {author} {\bibfnamefont {M.}~\bibnamefont
  {Woerner}}, \bibinfo {author} {\bibfnamefont {W.}~\bibnamefont {Kuehn}},
  \bibinfo {author} {\bibfnamefont {P.}~\bibnamefont {Bowlan}}, \bibinfo
  {author} {\bibfnamefont {K.}~\bibnamefont {Reimann}},\ and\ \bibinfo {author}
  {\bibfnamefont {T.}~\bibnamefont {Elsaesser}},\ }\href
  {https://doi.org/10.1088/1367-2630/15/2/025039} {\bibfield  {journal}
  {\bibinfo  {journal} {New J. Phys.}\ }\textbf {\bibinfo {volume} {15}},\
  \bibinfo {pages} {025039} (\bibinfo {year} {2013})}\BibitemShut {NoStop}%
\bibitem [{\citenamefont {Lu}\ \emph {et~al.}(2016)\citenamefont {Lu},
  \citenamefont {Zhang}, \citenamefont {Hwang~Harold}, \citenamefont
  {Ofori-Okai~Benjamin}, \citenamefont {Fleischer},\ and\ \citenamefont
  {Nelson~Keith}}]{Lu2016}%
  \BibitemOpen
  \bibfield  {author} {\bibinfo {author} {\bibfnamefont {J.}~\bibnamefont
  {Lu}}, \bibinfo {author} {\bibfnamefont {Y.}~\bibnamefont {Zhang}}, \bibinfo
  {author} {\bibfnamefont {Y.}~\bibnamefont {Hwang~Harold}}, \bibinfo {author}
  {\bibfnamefont {K.}~\bibnamefont {Ofori-Okai~Benjamin}}, \bibinfo {author}
  {\bibfnamefont {S.}~\bibnamefont {Fleischer}},\ and\ \bibinfo {author}
  {\bibfnamefont {A.}~\bibnamefont {Nelson~Keith}},\ }\href
  {https://doi.org/10.1073/pnas.1609558113} {\bibfield  {journal} {\bibinfo
  {journal} {Proc. Natl. Acad. Sci. U.S.A.}\ }\textbf {\bibinfo {volume}
  {113}},\ \bibinfo {pages} {11800} (\bibinfo {year} {2016})}\BibitemShut
  {NoStop}%
\bibitem [{\citenamefont {Chen}\ \emph {et~al.}(2020)\citenamefont {Chen},
  \citenamefont {Ye}, \citenamefont {Zhang},\ and\ \citenamefont
  {Li}}]{ChenYuYuan2020}%
  \BibitemOpen
  \bibfield  {author} {\bibinfo {author} {\bibfnamefont {Y.-Y.}\ \bibnamefont
  {Chen}}, \bibinfo {author} {\bibfnamefont {C.}~\bibnamefont {Ye}}, \bibinfo
  {author} {\bibfnamefont {Q.}~\bibnamefont {Zhang}},\ and\ \bibinfo {author}
  {\bibfnamefont {Y.}~\bibnamefont {Li}},\ }\href
  {https://doi.org/10.1063/5.0008157} {\bibfield  {journal} {\bibinfo
  {journal} {J. Chem. Phys.}\ }\textbf {\bibinfo {volume} {152}},\ \bibinfo
  {pages} {204305} (\bibinfo {year} {2020})}\BibitemShut {NoStop}%
\bibitem [{\citenamefont {Eibenberger}\ \emph {et~al.}(2017)\citenamefont
  {Eibenberger}, \citenamefont {Doyle},\ and\ \citenamefont
  {Patterson}}]{Eibenberger2017}%
  \BibitemOpen
  \bibfield  {author} {\bibinfo {author} {\bibfnamefont {S.}~\bibnamefont
  {Eibenberger}}, \bibinfo {author} {\bibfnamefont {J.}~\bibnamefont {Doyle}},\
  and\ \bibinfo {author} {\bibfnamefont {D.}~\bibnamefont {Patterson}},\ }\href
  {https://doi.org/10.1103/PhysRevLett.118.123002} {\bibfield  {journal}
  {\bibinfo  {journal} {Phys. Rev. Lett.}\ }\textbf {\bibinfo {volume} {118}},\
  \bibinfo {pages} {123002} (\bibinfo {year} {2017})}\BibitemShut {NoStop}%
\bibitem [{\citenamefont {Mukamel}(1995)}]{mukamel1995}%
  \BibitemOpen
  \bibfield  {author} {\bibinfo {author} {\bibfnamefont {S.}~\bibnamefont
  {Mukamel}},\ }\href@noop {} {\emph {\bibinfo {title} {Principles of Nonlinear
  Optical Spectroscopy}}},\ Oxford series in optical and imaging sciences\
  (\bibinfo  {publisher} {Oxford University Press},\ \bibinfo {year}
  {1995})\BibitemShut {NoStop}%
\bibitem [{\citenamefont {Cho}(2009)}]{Cho2009}%
  \BibitemOpen
  \bibfield  {author} {\bibinfo {author} {\bibfnamefont {M.}~\bibnamefont
  {Cho}},\ }\href@noop {} {\emph {\bibinfo {title} {Two-dimensional optical
  spectroscopy}}}\ (\bibinfo  {publisher} {CRC Press},\ \bibinfo {year}
  {2009})\BibitemShut {NoStop}%
\bibitem [{\citenamefont {Tian}\ \emph {et~al.}(2003)\citenamefont {Tian},
  \citenamefont {Keusters}, \citenamefont {Suzaki},\ and\ \citenamefont
  {Warren}}]{Tian2003}%
  \BibitemOpen
  \bibfield  {author} {\bibinfo {author} {\bibfnamefont {P.}~\bibnamefont
  {Tian}}, \bibinfo {author} {\bibfnamefont {D.}~\bibnamefont {Keusters}},
  \bibinfo {author} {\bibfnamefont {Y.}~\bibnamefont {Suzaki}},\ and\ \bibinfo
  {author} {\bibfnamefont {W.~S.}\ \bibnamefont {Warren}},\ }\href
  {https://doi.org/10.1126/science.1083433} {\bibfield  {journal} {\bibinfo
  {journal} {Science}\ }\textbf {\bibinfo {volume} {300}},\ \bibinfo {pages}
  {1553} (\bibinfo {year} {2003})}\BibitemShut {NoStop}%
\bibitem [{\citenamefont {Schlau-Cohen}\ \emph {et~al.}(2011)\citenamefont
  {Schlau-Cohen}, \citenamefont {Ishizaki},\ and\ \citenamefont
  {Fleming}}]{Fleming-CP2011}%
  \BibitemOpen
  \bibfield  {author} {\bibinfo {author} {\bibfnamefont {G.~S.}\ \bibnamefont
  {Schlau-Cohen}}, \bibinfo {author} {\bibfnamefont {A.}~\bibnamefont
  {Ishizaki}},\ and\ \bibinfo {author} {\bibfnamefont {G.~R.}\ \bibnamefont
  {Fleming}},\ }\href
  {https://doi.org/https://doi.org/10.1016/j.chemphys.2011.04.025} {\bibfield
  {journal} {\bibinfo  {journal} {Chem. Phys.}\ }\textbf {\bibinfo {volume}
  {386}},\ \bibinfo {pages} {1} (\bibinfo {year} {2011})}\BibitemShut {NoStop}%
\bibitem [{\citenamefont {Zare}(1988)}]{Zare1988}%
  \BibitemOpen
  \bibfield  {author} {\bibinfo {author} {\bibfnamefont {R.~N.}\ \bibnamefont
  {Zare}},\ }\href@noop {} {\emph {\bibinfo {title} {Angular Momentum:
  Understanding Spatial Aspects in Chemistry and Physics}}}\ (\bibinfo
  {publisher} {John Wiley},\ \bibinfo {year} {1988})\BibitemShut {NoStop}%
\bibitem [{\citenamefont {Lovas}\ \emph {et~al.}(2009)\citenamefont {Lovas},
  \citenamefont {Plusquellic}, \citenamefont {Pate}, \citenamefont {Neill},
  \citenamefont {Muckle},\ and\ \citenamefont {Remijan}}]{LOVAS2009}%
  \BibitemOpen
  \bibfield  {author} {\bibinfo {author} {\bibfnamefont {F.}~\bibnamefont
  {Lovas}}, \bibinfo {author} {\bibfnamefont {D.}~\bibnamefont {Plusquellic}},
  \bibinfo {author} {\bibfnamefont {B.~H.}\ \bibnamefont {Pate}}, \bibinfo
  {author} {\bibfnamefont {J.~L.}\ \bibnamefont {Neill}}, \bibinfo {author}
  {\bibfnamefont {M.~T.}\ \bibnamefont {Muckle}},\ and\ \bibinfo {author}
  {\bibfnamefont {A.~J.}\ \bibnamefont {Remijan}},\ }\href
  {https://doi.org/https://doi.org/10.1016/j.jms.2009.06.013} {\bibfield
  {journal} {\bibinfo  {journal} {J. Mol. Spectrosc.}\ }\textbf {\bibinfo
  {volume} {257}},\ \bibinfo {pages} {82} (\bibinfo {year} {2009})}\BibitemShut
  {NoStop}%
\bibitem [{\citenamefont {Arenas}\ \emph {et~al.}(2017)\citenamefont {Arenas},
  \citenamefont {Gruet},\ and\ \citenamefont {L.}}]{ARENAS2017}%
  \BibitemOpen
  \bibfield  {author} {\bibinfo {author} {\bibfnamefont {B.~E.}\ \bibnamefont
  {Arenas}}, \bibinfo {author} {\bibfnamefont {S.}~\bibnamefont {Gruet}},\ and\
  \bibinfo {author} {\bibfnamefont {A.}~\bibnamefont {L.}},\ }\href
  {https://doi.org/https://doi.org/10.1016/j.jms.2017.02.017} {\bibfield
  {journal} {\bibinfo  {journal} {J. Mol. Spectrosc.}\ }\textbf {\bibinfo
  {volume} {337}},\ \bibinfo {pages} {9} (\bibinfo {year} {2017})}\BibitemShut
  {NoStop}%
\bibitem [{\citenamefont {Patterson}\ and\ \citenamefont
  {Doyle}(2012)}]{Patterson2012}%
  \BibitemOpen
  \bibfield  {author} {\bibinfo {author} {\bibfnamefont {D.}~\bibnamefont
  {Patterson}}\ and\ \bibinfo {author} {\bibfnamefont {J.~M.}\ \bibnamefont
  {Doyle}},\ }\href {https://doi.org/10.1080/00268976.2012.679632} {\bibfield
  {journal} {\bibinfo  {journal} {Mol. Phys.}\ }\textbf {\bibinfo {volume}
  {110}},\ \bibinfo {pages} {1757} (\bibinfo {year} {2012})}\BibitemShut
  {NoStop}%
\bibitem [{\citenamefont {Zhang}\ and\ \citenamefont
  {Dong}(2022)}]{ZhangXue2022}%
  \BibitemOpen
  \bibfield  {author} {\bibinfo {author} {\bibfnamefont {X.}~\bibnamefont
  {Zhang}}\ and\ \bibinfo {author} {\bibfnamefont {H.}~\bibnamefont {Dong}},\
  }\href {https://doi.org/10.1103/PhysRevA.106.043516} {\bibfield  {journal}
  {\bibinfo  {journal} {Phys. Rev. A}\ }\textbf {\bibinfo {volume} {106}},\
  \bibinfo {pages} {043516} (\bibinfo {year} {2022})}\BibitemShut {NoStop}%
\end{thebibliography}%

\end{document}